\documentclass[aps,twocolumn,superscriptaddress,a4paper,prd]{revtex4-2}
\usepackage[colorlinks=true, pdfstartview=FitV, linkcolor=blue, citecolor=red, urlcolor=black]{hyperref}
\usepackage{graphicx}
\usepackage[all]{xy}
\usepackage{amsmath}
\usepackage{amssymb}
\usepackage{comment}
\usepackage{tensor}
\usepackage{orcidlink}
\newcommand{\be}{\begin{equation}}
\newcommand{\ee}{\end{equation}}
\newcommand{\ben}{\begin{eqnarray}}
\newcommand{\een}{\end{eqnarray}}
\newcommand{\bes}{\begin{subequations}}
\newcommand{\ees}{\end{subequations}}
\def\bal#1\eal{\begin{align}#1\end{align}}

\newcommand{\bfi}{\begin{figure}}
\newcommand{\efi}{\end{figure}}
\newcommand{\bc}{\begin{center}}
\newcommand{\ec}{\end{center}}

\newcommand{\csch}{\mbox{csch}}
\newcommand{\sech}{\mbox{sech}}
\newcommand{\arcsinh}{\mbox{arcsinh}}

\newcommand{\LL}{{\cal L}}

\begin{document}
\title{Large lumps}
\author{D. Bazeia\,\orcidlink{0000-0003-1335-3705}}
     \email[]{dbazeia@gmail.com}
    \affiliation{Departamento de F\'\i sica, Universidade Federal da Para\'\i ba, 58051-970 Jo\~ao Pessoa, PB, Brazil}
\author{I. Bezerra\,\orcidlink{0009-0000-6754-5943}}
     \email[]{irb@academico.ufpb.br}
     \affiliation{Departamento de F\'\i sica, Universidade Federal da Para\'\i ba, 58051-970 Jo\~ao Pessoa, PB, Brazil}
\author{M.A. Marques\,\orcidlink{0000-0001-7022-5502}}
        \email[]{mam@fisica.ufpb.br}
    \affiliation{Departamento de F\'\i sica, Universidade Federal da Para\'\i ba, 58051-970 Jo\~ao Pessoa, PB, Brazil}
\author{R. Menezes\,\orcidlink{0000-0002-9586-4308}}
     \email[]{rmenezes@dcx.ufpb.br}\affiliation{Departamento de Ci\^encias Exatas, Universidade Federal da Para\'{\i}ba, 58297-000 Rio Tinto, PB, Brazil}
\begin{abstract}
We introduce a procedure to obtain lump solutions via the formation of a kink-antikink pair, consisting of the superposition of kinks whose distance from the origin is controlled by a single parameter $a$. For large values of $a$, a wide plateau appears in the solution, which we call a large lump. The procedure involves the use of a first-order equation that allows the construction of the potential associated with the lump solution. We then investigate several known scalar field models where the parent kinks are capable of giving rise to novel lumps. The lump inherits the tails of the parent kink, allowing for either short-range exponential profiles or long-range profiles characterized by distinct power-law decays. We also show how to verify if an arbitrary lump solution can be obtained via our method and illustrate this possibility with a novel vacuumless lump. 
\end{abstract}

\maketitle

\section{Introduction}
Scalar field models are of interest in Field Theory, particularly in the investigation of localized structures. In the presence of nonlinear interactions, the equations of motion may support static configurations with finite energy, which can be of topological or non-topological nature \cite{manton}. The simplest topological structures are kinks, which appear in models of a single real scalar field in $(1,1)$ spacetime dimensions and connect two neighboring minima of the potential \cite{vachaspati,Shnir}. They are stable, minimum-energy solutions. In $(1,1)$ dimensions, the non-topological solutions that also arise in scalar field models are usually called lumps \cite{Shnir,wilets,lump1}; they depart from and return to the same minimum of the potential. They are not minimum-energy configurations and are unstable under small fluctuations. Even so, they may be useful in applications involving high energy physics \cite{RPP}, optical communications \cite{optics1,optics2,OS}, Bose-Einstein condensates \cite{BE1,BE2,BE3,BE4}, Q-balls \cite{coleman,RPP}, and braneworlds \cite{brane}. In particular, lumps in scalar field models are directly related to the presence of bright solitons in fibers \cite{BSf} and matter-wave bright solitons in condensates \cite{BEb}, guided by specific interactions in the nonlinear Schrödinger equation or the Gross-Pitaevskii equation; see, e.g., Ref. \cite{Malomed} for the case that includes a nonlinear optical lattice.

The lump is perhaps the simplest case of a sphaleron, a term coined in Ref.~\cite{sphaleron1} to name a static, unstable, finite-energy solution to the classical field equations of the electroweak sector in the Standard Model of particle physics. Over the years, several papers have investigated such lumplike configurations \cite{sphaleron2,sphaleron3,sphaleron4,sphaleron5,sphaleron6,sphaleron7,sphaleron8,sphaleron9,sphaleron10}. In particular, a topic of current interest is the relationship between the emergence of sphalerons and oscillons, although they are distinct non-perturbative objects. Another topic directly connected to lumps in scalar field models concerns the presence of bell-shaped soliton solutions for the Korteweg-de Vries and other nonlinear equations \cite{Whi,Das1}. In this context, one notes that Ref. \cite{Das2} includes an algebraic method for generating classes of traveling wave solutions for a variety of partial diﬀerential equations of current interest in nonlinear science.  In this sense, the procedure to be developed below may also be of interest to the subject of integrable systems \cite{Das1,Das2,Outro}. 

If one pays attention to the search for solutions to scalar field models, the properties of localized structures are related to the way the potential behaves near its minima. In the case of lumps, the second derivative of the potential at the minimum related to the solution controls the classical mass and determines whether the tails of the lump are short range, with exponential falloff, or long range, with power-law behavior. This point is important because the asymptotic profile of the solutions controls the interaction between the structures. Over the years, several mechanisms have been proposed to modify the core or the tail of localized solutions, leading to structures with internal plateaus, compact or half-compact support, long-range behavior and other nonstandard profiles \cite{avelar,complump,lumpmarques,lumpscatt,lumpepl,XX1,XX2,XX3,XX4}. These results motivate the search for analytical procedures that allow one to obtain novel lump configurations with controllable shape and asymptotic behavior.

In this work, we develop a procedure to construct lump solutions from kink and antikink configurations. The idea is to start from a parent model that supports a kink in a given topological sector and use the superposition of a shifted kink and a shifted antikink to generate a non-topological solution. The distance of each substructure to the origin is controlled by a parameter $a$. As $a$ increases, the kink and antikink become more separated, and the resulting lump develops a wide plateau around the origin. Due to this feature, we call these configurations large lumps, since we can consider large values of the parameter $a$.

In order to implement the above procedure, we organize the work as follows. In Sec.~\ref{sec2}, the methodology is introduced on general grounds, and in Secs.~\ref{secsymmetric} and \ref{secasymmetric}, we apply the method to symmetric and asymmetric kinks, respectively. In Sec.~\ref{model4}, we discuss a special situation and present a vacuumless example. We then end the work in Sec.~\ref{secfinalremarks}, summarizing our results and adding perspectives for future investigations.

\section{Methodology} \label{sec2}
Let us consider the action of a single real scalar field in $(1,1)$ spacetime dimensions, $S=\int dx\,dt\,\LL$, with the Lagrangian density
\begin{equation} \label{lmodel}
    \mathcal{L} = \frac{1}{2}\partial_\mu\phi\partial^\mu\phi - V(\phi).
\end{equation}
We are using natural units, with dimensionless field and coordinates. We are interested in studying defect structures, so we investigate static configurations, $\phi=\phi(x)$, whose equation of motion is
\begin{equation}\label{eom}
 \phi'' = V_\phi.
\end{equation}
The prime stands for derivative with respect to $x$, so $\phi''=d^2\phi/dx^2$ and $V_\phi=dV/d\phi$. As is known in the literature \cite{vachaspati}, the above equation may support, among other possibilities, two important types of localized solutions: topological, such as kinks, and non-topological, such as lumps. Kinks are monotonic and connect two neighboring degenerate minima of the potential, obeying $\phi_k(\pm\infty)=v_\pm$ and $\phi'_k(\pm\infty)=0$, where $v_\pm$ are neighboring minima of the potential. Lumps are non-topological localized solutions satisfying $\phi_l(\pm\infty)=v$, where $v$ is a minimum of the potential. They reach a turning point at $x=x_0$, where $\phi_l(x_0)=\phi_0$ and $V(\phi_0)=0$, with $\phi_0$ not being a minimum of the potential. For the sake of clarity, we shall refer to potentials supporting kinks as $V_k(\phi)$ and the ones supporting lumps as $V_l(\phi)$. Also, without loss of generality, we take $x_0=0$ to get the center of the lump at the origin.

\begin{figure}[t!]
    \centering
    \includegraphics[width=0.75\linewidth]{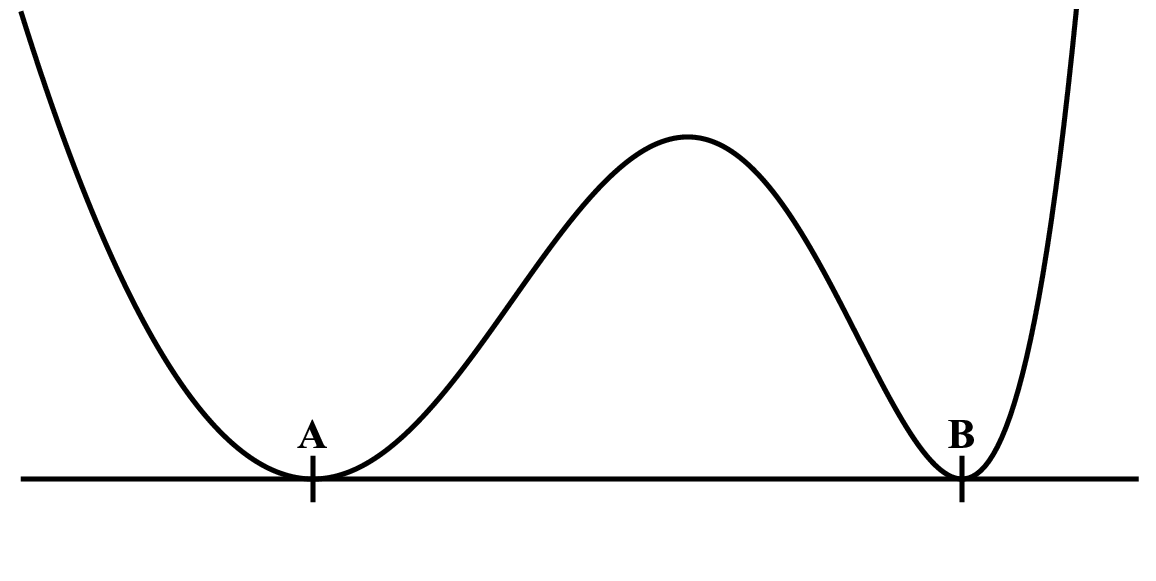}
    \includegraphics[width=0.75\linewidth]{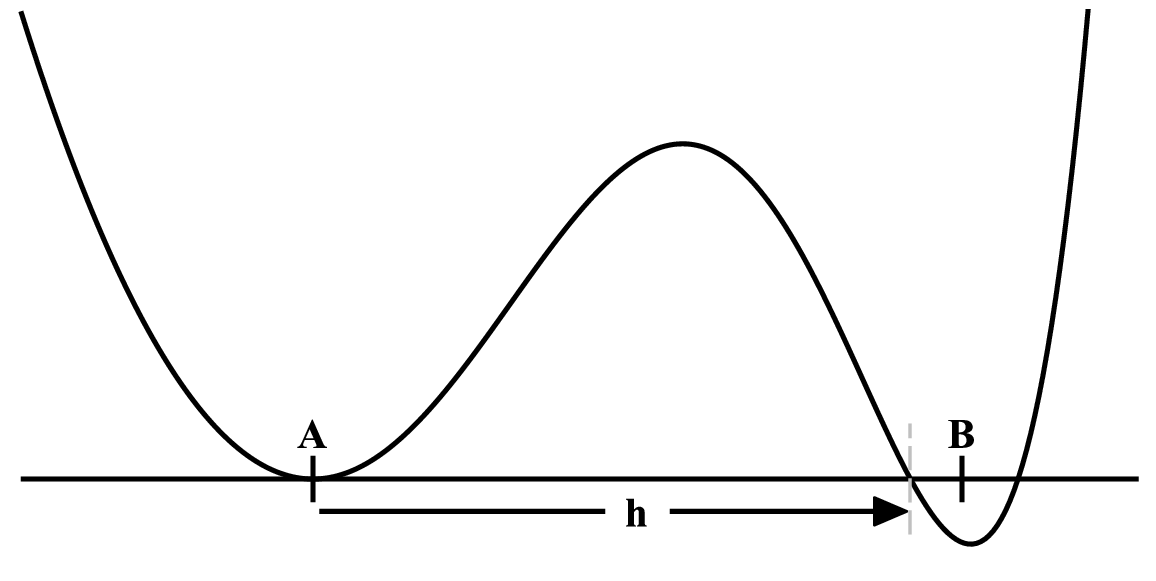}
    \includegraphics[width=0.75\linewidth]{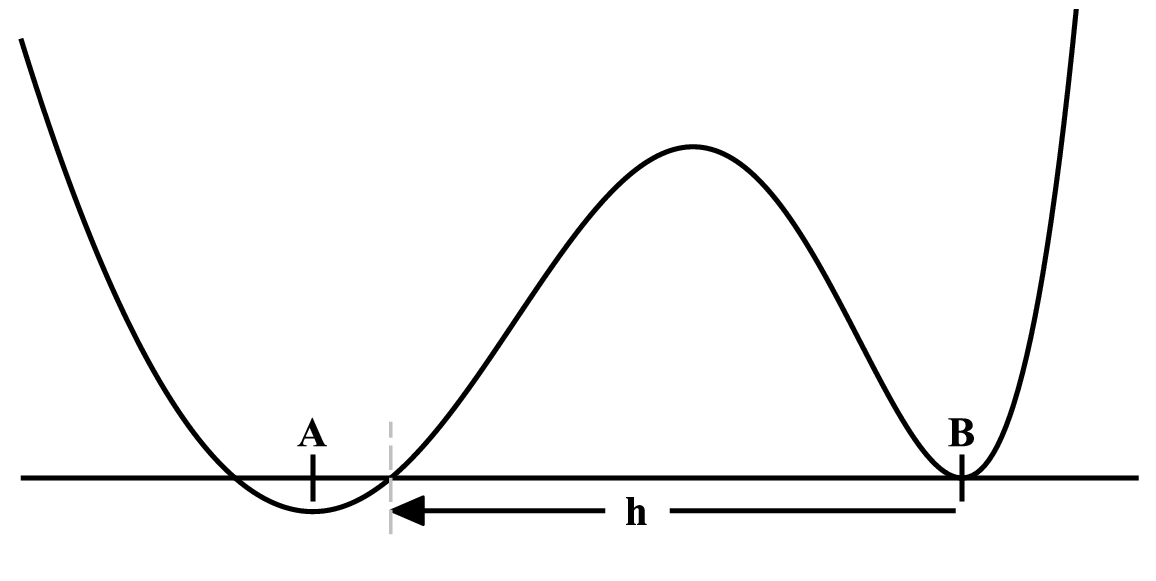}
    \caption{A pictorial illustration of the potential $V_k(\phi)$ (top), $V_l^a(\phi)$ with the right minimum shifted down (middle) and $V_l^a(\phi)$ with the left minimum shifted down (bottom).}
    \label{figesquema}
\end{figure}

As is well known, invariance under spacetime translations of \eqref{lmodel} is related to the presence of the energy-momentum tensor, which leads to the energy density $\rho=-\LL$ in the static case, which can be written in the form
\begin{equation}\label{rho}
    \rho(x) = \frac{1}{2} {\phi'}^2 + V(\phi).
\end{equation}
The above equation allows us to show that, to get finite-energy solutions, one may integrate the equation of motion \eqref{eom} into the following first-order equation
\be\label{fo}
\frac12{\phi'}^2=V(\phi).
\ee
In Ref.~\cite{avelar}, it was shown that a lump solution related to a $\phi^4$ potential can be obtained from a linear combination of kinks. We get inspiration from this solution and attempt to generalize it to the superposition of arbitrary kinks. So, let us suppose that some non-negative potential $V_k(\phi)$ supports at least one topological sector. In the top panel of Fig.~\ref{figesquema}, we illustrate it with the interval $[A,B]$, with $V_k(A)=V_k(B)=0$. In this sector, the solution which connects the minima $\phi=A$ to $\phi=B$ as $x$ spans from $-\infty$ to $\infty$ is a kink, which we call $\phi_k(x)$. Due to the invariance of the equation of motion \eqref{eom} under spatial reflections, $x\to-x$, the aforementioned topological sector also supports an antikink, $\phi_{ak}$, such that $\phi_{ak}(x)=\phi_k(-x)$. Even though we have defined the kink/antikink as the increasing/decreasing solution, one may consider the other way around.

Starting with the potential $V_k(\phi)$, we can construct a potential $V_l^a(\phi)$ by bringing a minimum of $V_k(\phi)$ down via the action of a parameter $a$, which is introduced to make the potential obey $\lim_{a\to\infty} V^a_l(\phi)=V_k(\phi)$. By doing so, the topological sector $[A,B]$ is lost, and the solution becomes a lump, denoted by $\phi_l(x)$, departing from the non-shifted minimum, passing through a non-minimum zero (point of return) of the potential $V^a_l(\phi)$ and returning to the aforementioned minimum.

Notice that there may be two possibilities for each topological sector of the potential $V_k(\phi)$, as one may shift down the left or right minimum; they are depicted in the middle and bottom panels of Fig.~\ref{figesquema}. The non-topological solution $\phi_l(x)$ which should appear in the middle (bottom) panel departs from the minimum $\phi=A$ ($\phi=B$) at $x\to-\infty$, increases (decreases) until the point of return at $x=0$, and goes back to the aforementioned minimum as $x\to\infty$. Since in our construction (see below) the limit $a\to\infty$ recovers the kink potential, the solution can be seen as a kink-antikink pair separated around the origin by a distance $\ell$ which depends on $a$; $\ell=\ell(a)$. In this situation, the point of return tends to $\phi=B$ ($\phi=A$), with a large plateau, such that the height approaches $h=|B-A|$.

The lump configurations can be obtained by combining kinks and antikinks in the most obvious manner, which consists of the superposition of a shifted kink and a shifted antikink, separated by some distance. For a given topological sector of the potential $V_k(\phi)$, there are two possibilities, as we have previously commented. They are
\bes\label{lumpgeral}
\bal\label{solmais}
&\phi^+_l(x)=\phi_k(x+a)+\phi_{ak}(x-a)-\phi_k(\infty),\\ \label{solmenos}
&\phi^-_l(x)=\phi_{ak}(x+a) + \phi_k(x-a) - \phi_{ak}(\infty),
\eal
\ees
which are centered at $x=0$. Asymptotically, we see that $\phi^+_l(\pm\infty) = \phi_k(-\infty) = A$ and $\phi^-_l(\pm\infty) = \phi_k(+\infty) = B$.  One must be careful with the above expressions for $\phi^\pm_l$, as one cannot be sure that they will always lead to acceptable lumps because they may engender critical points outside the origin. Therefore, one must investigate the presence of maxima and minima outside the point of return ($x=0$), because such a behavior is incompatible with an acceptable potential. If an inflection point arises, however, the lump solution is compatible with the equations of motion. Since the lumps constructed with our method have a point of return at $x=0$, we have $\phi_l^+(0) = 2\phi_k(a) - B$ and $\phi_l^-(0)=2\phi_k(-a)-A$. In the limit $a\to\infty$, we get $\phi_l^+(0)=B$ and $\phi_l^-(0)=A$.

Given the above solutions, one can construct two potentials, $V_l^{a\pm}(\phi)$, using Eq.~\eqref{fo}, with the plus/minus superscript representing the case in which the right/left minimum of $V_k(\phi)$ is shifted. We remark, however, that the construction of the potential $V_l^{a\pm}(\phi)$ is not always feasible analytically, as the above solutions are formed by the superposition of non-linear functions. Indeed, by looking at both expressions in \eqref{lumpgeral}, we see that the nonlinearity introduces an obstacle to inverting $\phi_l^\pm(x)$ and writing $x$ as a function of $\phi$, which is essential to obtain the exact expression of the potential $V^\pm_l(\phi)$.

It is worth commenting that the resulting model obtained by the superposition of kinks depends on their symmetry. Let us consider a symmetric kink connecting the points $\phi=A$ and $\phi=B$, obeying $\phi_{ak}(x)=A+B-\phi_k(x)$, so the potential $V_k(\phi)$ supports reflection symmetry around the point $\phi=(A+B)/2$. In this situation, the case in which $V_l(\phi)$ is formed by shifting down $\phi=A$ is equivalent to the one constructed from the displacement of $\phi=B$, therefore a single model can be constructed. Thus, without loss of generality, we only investigate one of the cases. On the other hand, asymmetric kinks may lead to two distinct models; this case is illustrated in Sec.~\ref{secasymmetric}.

The asymptotic behavior of the lump inherits the tails of the kink and antikink used in the superposition. If we consider the $\phi_l^+(x)$ solution, we can define $\delta^+(x)=\phi^+_l(x)-A$ for $x\to\pm\infty$, $\delta_R(x)=\phi_k(x)-B$ for $x\to\infty$ and $\delta_L(x)=\phi_k(x)-A$ for $x\to-\infty$ to get, from Eq.~\eqref{solmais},
\be\label{deltamais}
\delta^+(x) = \delta_L(a-|x|) + \delta_R(a+|x|).
\ee
We can also define $\delta^-(x)=\phi^-_l(x)-B$ to show that the asymptotic behavior of the $\phi_l^-(x)$ solution in Eq.~\eqref{solmenos} is
\be\label{deltamenos}
\delta^-(x) = \delta_L(-|x|-a) + \delta_R(|x|-a).
\ee
Thus, the lump tail is controlled by the slowest decay among the two kink tails entering the superposition.

Let us first analyze the case of kinks and antikinks with exponential tails, i.e.,
\be\label{deltasexponential}
\delta_L(x) = C_Le^{m_L x}\quad\text{and}\quad \delta_R(x) = C_Re^{-m_R x},
\ee
where $m_L=\sqrt{V_{k\phi\phi}(A)}$ and $m_R=\sqrt{V_{k\phi\phi}(B)}$ are the classical masses of the minima related to the kink and antikink that form the lump and the constants $C_L$ and $C_R$ depend on the specific model of interest. By substituting this into \eqref{deltamais} and \eqref{deltamenos}, we see that the dominant term is exponential, with the form $\delta^\pm(x)\propto e^{-m_{\rm tail} |x|}$, where $m_{\rm tail}=\min(m_L,m_R)$. It is worth commenting that, for asymmetric kinks, a special situation may occur: the case of equal masses, $m=m_L=m_R$, and $C_L= -C_R e^{-2ma}$ for $\delta^+(x)$ and $C_R= -C_L e^{-2ma}$ for $\delta^-(x)$. These specific values require us to consider the next term in the asymptotic expansion of $\delta_L(x)$ and $\delta_R(x)$, so we shall have $\delta^\pm(x)\propto e^{-\mu|x|}$, where $\mu>m$.

Another possibility is the case of kinks with power-law tails, which appear for null classical masses, with
\be\label{deltaspowelaw}
\delta_L(x) = \frac{C_L}{(-x)^{n_L}}\quad\text{and}\quad \delta_R(x) =\frac{C_R}{x^{n_R}},
\ee
where $n_L$ and $n_R$ are positive real numbers that control the falloff at the left and right sides of the kink. This can be used in \eqref{deltamais} and \eqref{deltamenos} in order to determine the decay of the lump. Similarly to the previous case, the dominant term is the one with the slower falloff, so we have $\delta(x)\propto1/|x|^n$, where $n=\min(n_L,n_R)$. The case in which $n=n_L=n_R$ is also special; it may require us to consider more terms in $\delta_L(x)$ and $\delta_R(x)$ as we shall see in an example of this work. In particular, this situation arises for $C_L=-C_R$.

Of course, there are many other possibilities for the asymptotic behavior, such as the quasi-compact or vacuumless tails. Their general aspects are not discussed here.

\begin{figure}[t!]
    \centering
    \includegraphics[width=0.75\linewidth]{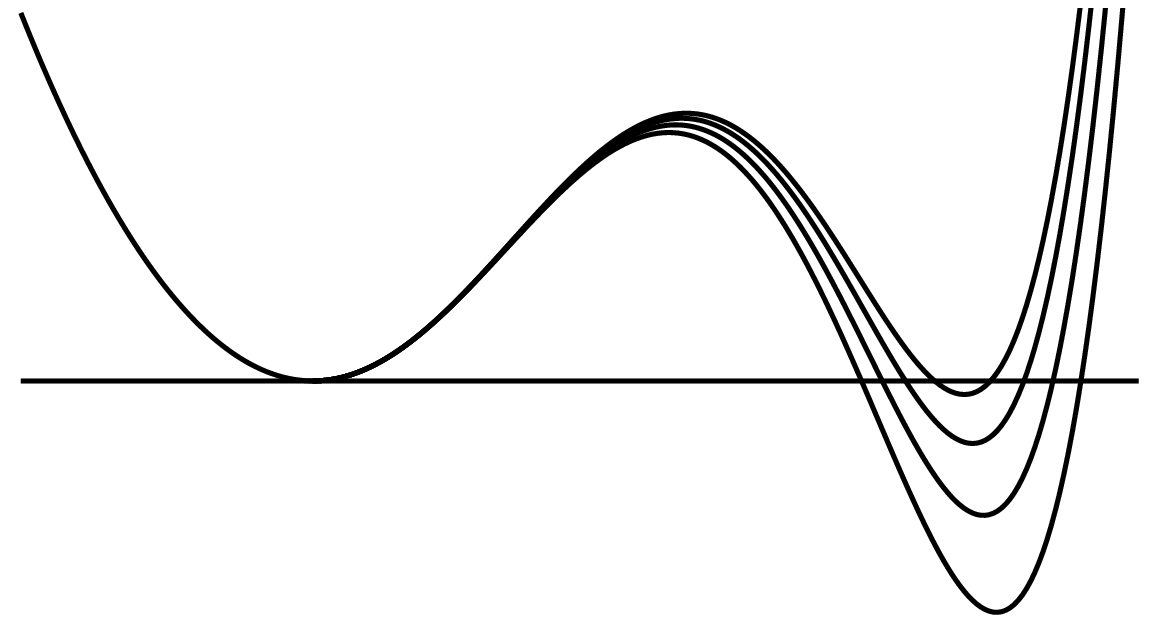}
    \caption{A pictorial illustration of the potential $V_l^a(\phi)$ showing that, as $a$ gets larger, the minimum on the right goes up and the slope of the potential at the zero (return point of $\phi_l(x)$) decreases.}
    \label{figesquema2}
\end{figure}

We can further analyze the behavior of the lump at the origin by expanding it, $\phi^\pm_l(x) \approx \phi^\pm_l(0) + \alpha_\pm x^2/2$, where $\alpha_\pm={V^{a\pm}_{l\phi}}(\phi^\pm_l(0))$. We display a generic situation in Fig.~\ref{figesquema2}, where we show that, as $a$ gets larger, the minimum on the right of the potential goes up and the return point of the lump solution leads to a smaller slope ($\alpha_\pm\to0$) in the potential. This means that $\phi^\pm_l(x)$ indeed gets a wider plateau around the origin as $a$ increases.

Considering the lump in the form \eqref{lumpgeral}, we can calculate the energy density \eqref{rho}, combined with the first-order equation \eqref{fo}, which leads to
\bes\label{rholump}
\bal\label{rhomais}
& \rho^+_l(x) =\rho_k(x+a) + \rho_{ak}(x-a) + 2{\phi'_k}(x+a)\,{\phi'_{ak}}(x-a),\\ \label{rhomenos}
& \rho^-_l(x) =\rho_{ak}(x+a) + \rho_{k}(x-a) + 2{\phi'_{ak}}(x+a)\,{\phi'_k}(x-a),
\eal
\ees
where $\rho_k(x)={\phi^{\prime2}_k}(x)$ and $\rho_{ak}(x)={\phi^{\prime2}_{ak}}(x)$.

By integrating the above expressions, we get the energy
\bes
\bal
E^{a-}_l &= 2E_k - 2\int_{-\infty}^{+\infty} dx\,|{\phi'_{ak}}(x+a)\,{\phi'_{k}}(x-a)|,\\
E^{a+}_l &= 2E_k - 2\int_{-\infty}^{+\infty} dx\,|{\phi'_k}(x+a)\,{\phi'_{ak}}(x-a)|,
\eal
\ees
where $E_k$ is the energy of the kink and we have used the fact that kink and antikink have slopes with opposite signs everywhere. As $a$ gets larger, the energy approaches $E^{a\pm}_l\approx 2E_k$.

As is well known, lumps are linearly unstable. Indeed, by considering small fluctuations $\eta(x,t)=\sum_k\eta_k(x)\cos(\omega_k t)$ around the static solutions, we get the Schr\"odinger-like stability equation $-\eta_k'' + U(x)\eta_k = \omega_k^2\eta_k $ with potential
\be\label{stabpot}
U(x) =\frac{\phi'''}{\phi'}=V_{\phi\phi}(\phi(x)),
\ee
where $\phi(x)$ stands for the static solutions of Eqs.~\eqref{lumpgeral}. This potential always leads to the presence of a negative eigenvalue, leading to instability.

In the next section, we illustrate how to construct models that support lump solutions \eqref{lumpgeral} using symmetric kinks with distinct features. 

\section{Superposition of symmetric kinks}\label{secsymmetric}
\begin{figure}
    \centering
    \includegraphics[width=0.75\linewidth]{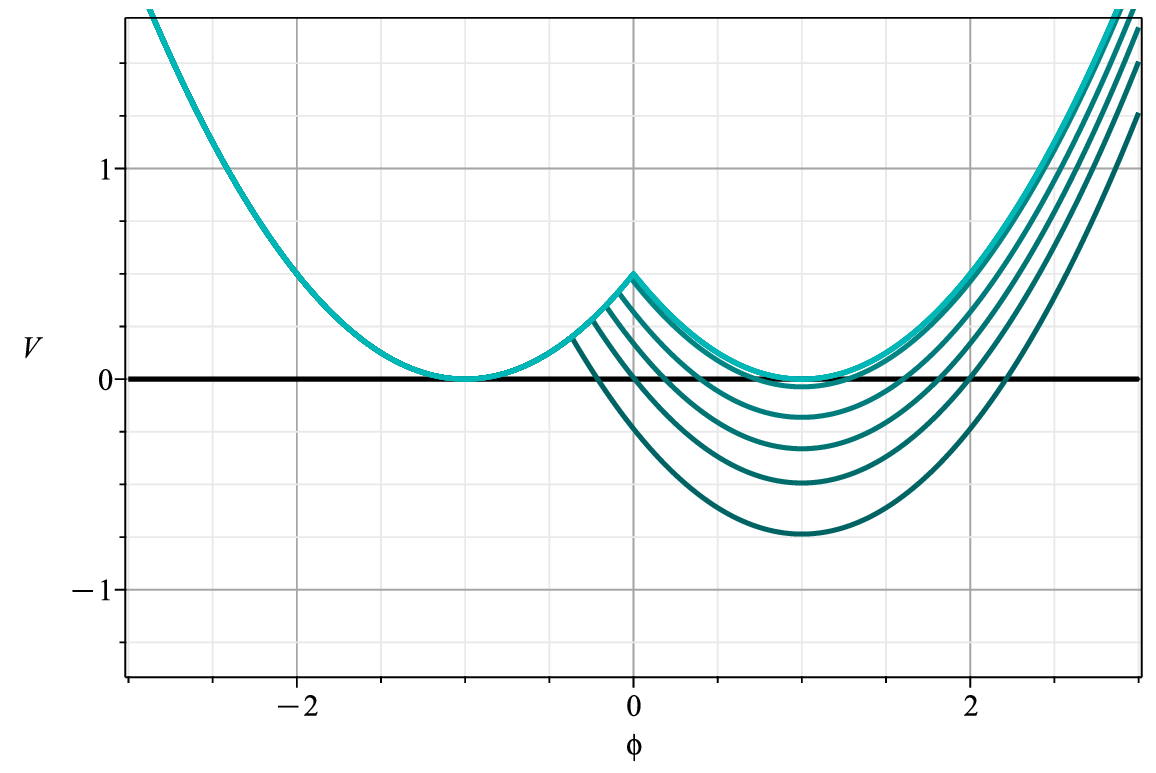}
    \caption{The potential \eqref{potlumptoy} depicted for $a=0.5,0.7,0.9,1.2,2,4,6,8$ and $10$. The colors get lighter as $a$ gets larger. The curves for $a\geq2$ become indistinguishable, appearing as the curve with the lightest color.}
    \label{fig:pottoy}
\end{figure}

Let us now illustrate our procedure. We first consider a toy model \cite{toy1,toy2} given by
\be
V_k^{\rm toy}(\phi) = \frac12(1-|\phi|)^2.
\ee
This potential has minima at $\phi=\pm1$ and a maximum at $\phi=0$. It supports the kink solution
\be
\phi^{\rm toy}_k(x)={\rm sgn}(x)(1-e^{-|x|}).
\ee
Notice that the above kink is actually obtained by gluing exponential tails that arise from potentials with finite classical masses at their minima. The energy density of the above kink is $\rho_k(x)=e^{-2|x|}$, which leads to the energy $E=1$. We can use Eq.~\eqref{solmais} to construct the lump solution
\be\label{lumptoy}
\phi_l(x) = \begin{cases}
     2e^{-|x|}\sinh(a)-1,&|x|>a\\
     1-2e^{-a}\cosh(x),&|x|\leq a.
\end{cases}
\ee
Its associated energy density can be written from Eq.~\eqref{rhomais}, which leads us to
\be\label{rholumptoy}
\rho_l(x) = \begin{cases}
     4e^{-2|x|}\sinh^2(a),&|x|>a\\
     4e^{-2a}\sinh^2(x),&|x|\leq a.
\end{cases}
\ee
\begin{figure}
    \centering
    \includegraphics[width=0.5\linewidth]{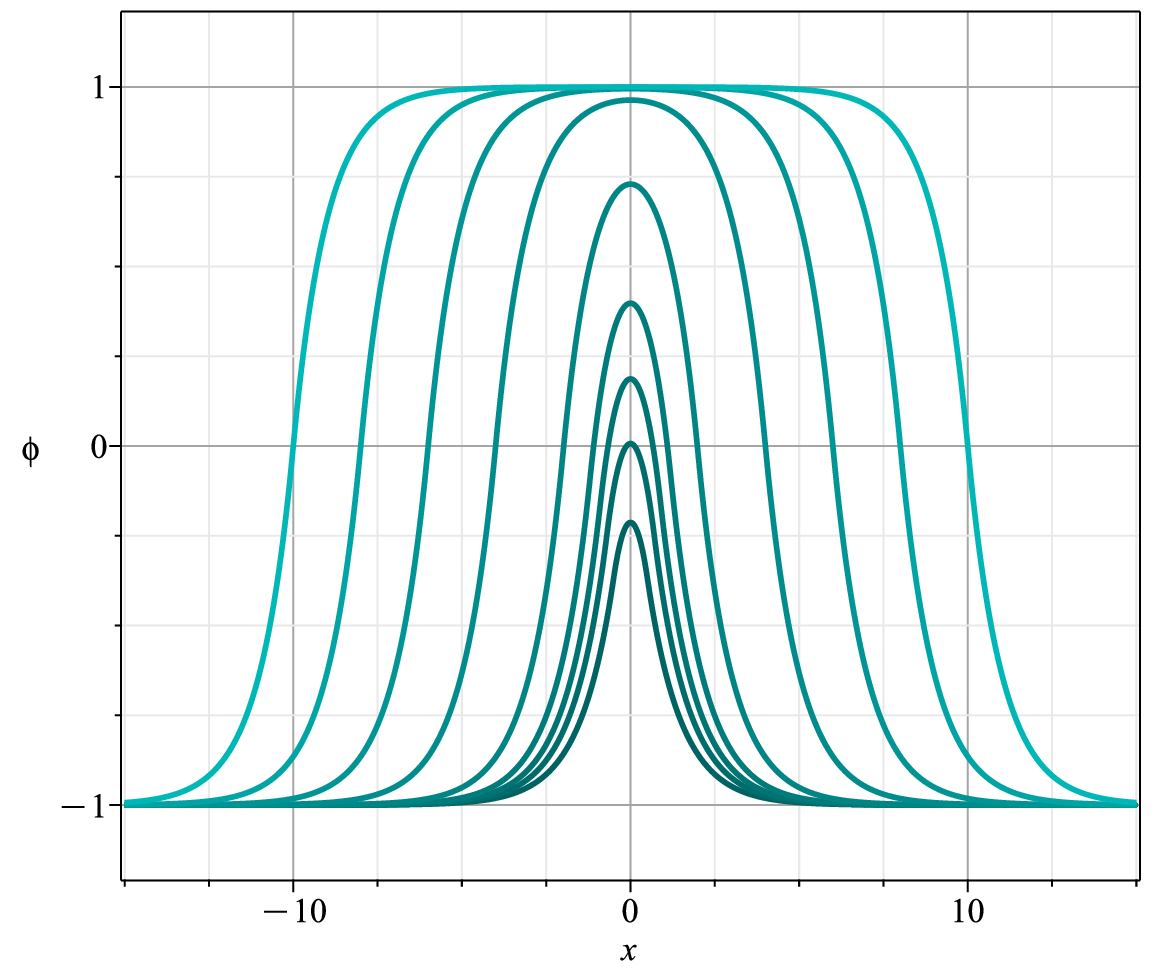}\includegraphics[width=0.5\linewidth]{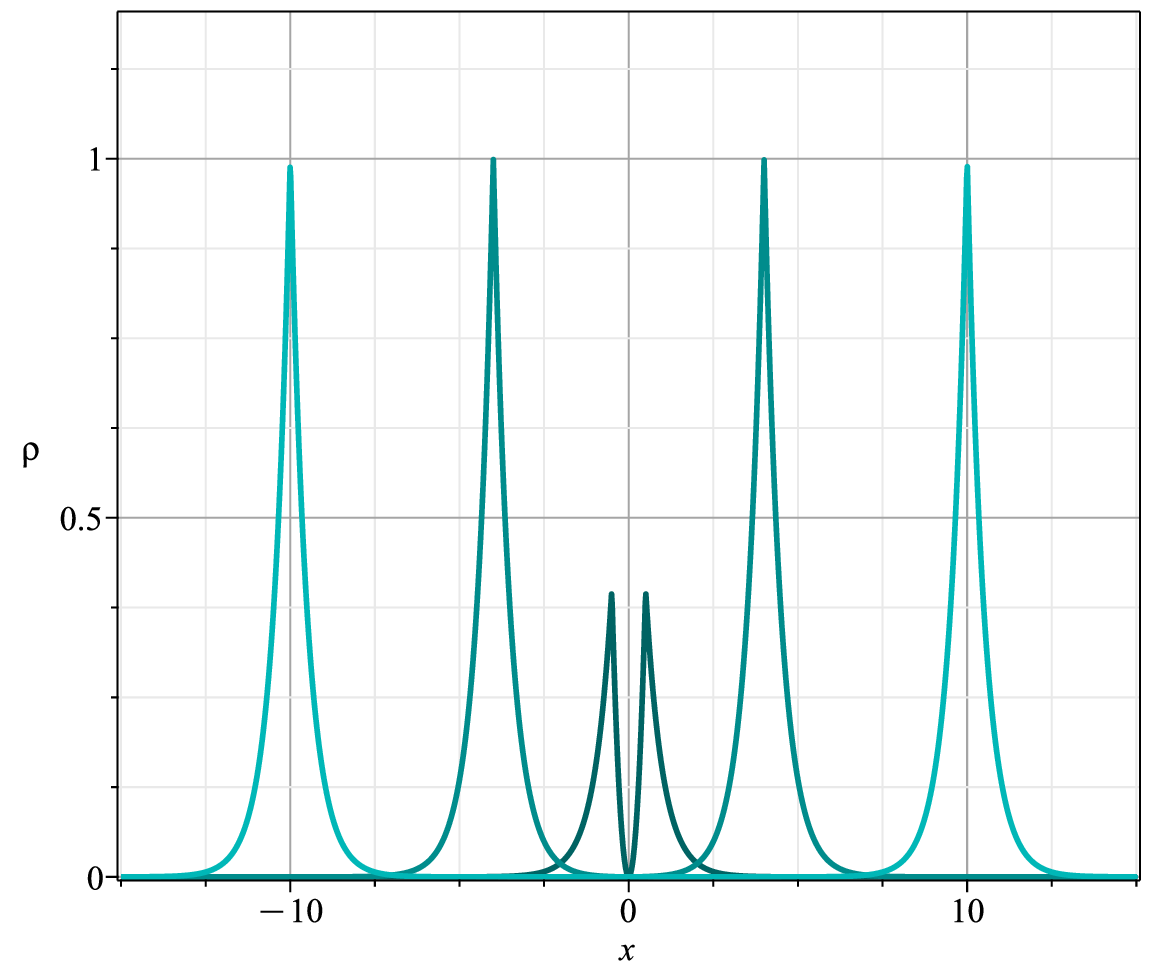}
    \caption{The solution \eqref{lumptoy} (left) depicted for $a=0.5,0.7,0.9,1.2,2,4,6,8$ and $10$ and the energy density \eqref{rholumptoy} (right), for $a=0.5, 4$ and $10$. The colors get lighter as $a$ gets larger.}
    \label{fig:lumptoy}
\end{figure}
By integrating it, we get the energy $E_l = 2-(2+4a)e^{-2a}$. In Fig.~\ref{fig:lumptoy} we depict the solution \eqref{lumptoy} and its energy density \eqref{rholumptoy}. Notice that the parameter $a$ controls the width of the lump. For $a\to\infty$, the solution is equivalent to two kinks separated by an infinite distance, each one accounts for $E_k=1$ in the energy; this explains the behavior $E_l\to2E_k=2$ for $a\to\infty$. The potential associated with the lump \eqref{lumptoy} is
\be\label{potlumptoy}
V_l(\phi) = \begin{cases}
    \frac12(1+\phi)^2,&\phi\leq-e^{-2a}\\
    \frac12(1-\phi)^2-2e^{-2a}, &\phi>-e^{-2a}.
\end{cases}
\ee
This potential has a minimum at $\phi=-1$ and a zero, which is the point of return of the solution, at $\phi=1-2e^{-a}$. As $a$ gets larger, this zero tends to the minimum $\phi=1$. In Fig.~\ref{fig:pottoy}, we depict the above potential for several values of $a$.

The linear stability is investigated via a Schr\"odinger-like equation with the stability potential and lowest eigenvalue respectively given by
\bal
& U(x)=1-\frac{2}{1-e^{-2a}}\left[\delta(x-a)+\delta(x+a)\right],\\
&\omega_{g}^2=1-\left[\frac{1}{1-e^{-2a}}+\frac{1}{2a}W\!\left(\frac{2a\,e^{-\frac{2a}{1-e^{-2a}}}}{1-e^{-2a}}\right)\right]^2.
\eal
In the latter expression, $W$ is the main branch of Lambert function. Since $\omega^2_g$, which is the eigenvalue of the ground state, is negative, the lump solution \eqref{lumptoy} is unstable, as expected. We also remark that all the other modes associated with the above stability potential can be investigated analytically.

Next, in this section, we show that a lump solution which emerges from a known $\phi^4$ model is compatible with our method and, later, we present novel lump solutions from sine-Gordon, double sine-Gordon and long-range sine-Gordon models.

\subsection{Lumps from $\phi^4$ model}
Let us construct a lump from the well-known $\phi^4$ model, with
\be\label{phi4modelkink}
V_k(\phi)=\frac12(1-\phi^2)^2,\quad \phi_k(x) = \tanh(x).
\ee
The above kink solution has the energy density $\rho_k(x) = \sech^4(x)$, which leads to the energy $E_k=4/3$. We then review the model introduced in Ref.~\cite{avelar} and consider a shifted version of it, described by the potential
\be\label{potphi4}
\begin{aligned}
	V^a_l(\phi) &= \frac12\left(1+\phi\right)^2\left(\phi+1-2\coth(a)\right)\\
	&\hspace{4mm}\cdot\left(\phi+1-2\tanh(a)\right),
\end{aligned}
\ee
for $a>0$. The shift in the potential was done to make sure that it has the exact form obtained with our method. It supports zeros at $\phi=-1$, $\phi= 2\tanh(a)-1$ and $\phi=2\coth(a)-1$. We remark that $\phi=-1$ is a minimum regardless of the value of $a$. There is a second minimum, at $\phi=(e^{4 a}+\sqrt{e^{8 a}+34 e^{4 a}+1}+5)/(2( e^{4 a}-1))$. The maximum is located at $\phi=(e^{4 a}-\sqrt{e^{8 a}+34e^{4a}+1}+5)/(2(e^{4 a}-1))$. As expected, the limit $a\to\infty$ leads to the well-known $\phi^4$ model in Eq.~\eqref{phi4modelkink}. The change $\phi\to-\phi$ leads to the reflected situation.
\begin{figure}
    \centering
    \includegraphics[width=0.75\linewidth]{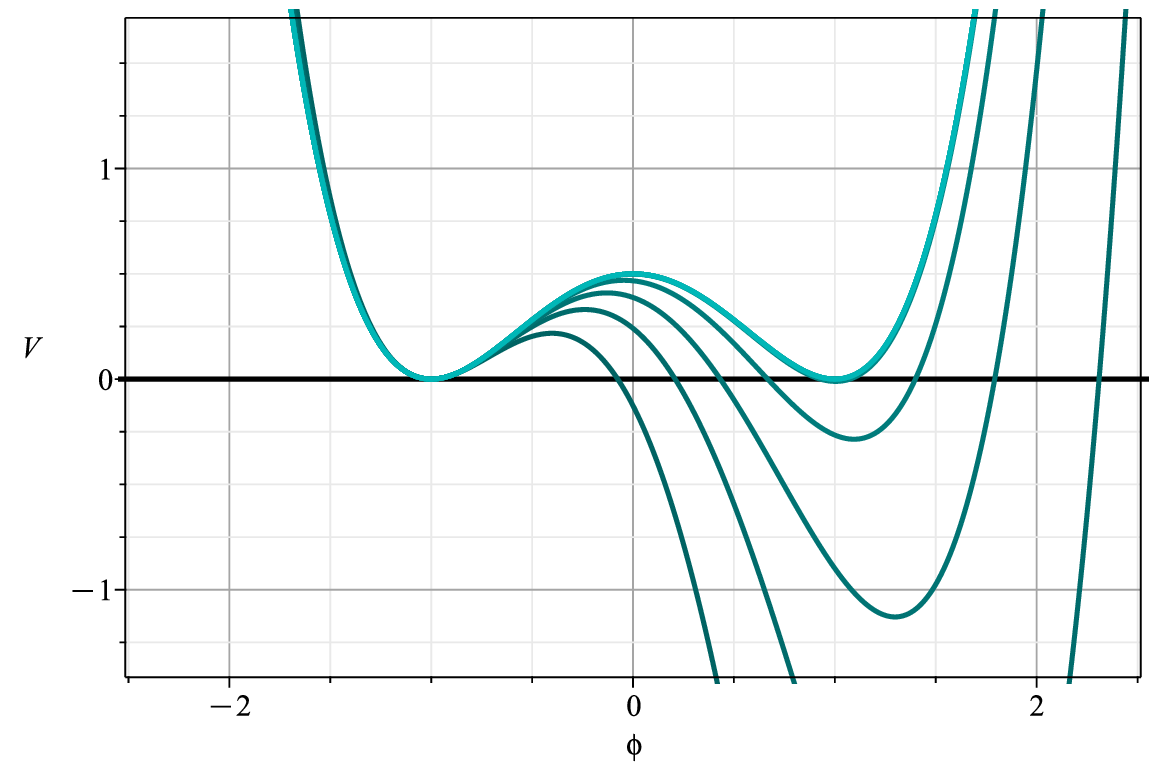}
    \caption{The potential \eqref{potphi4} depicted for $a=0.5,0.7,0.9,1.2,2,4,6,8$ and $10$. The shades of green follow the previous figures. For $a\geq2$, the lines become indistinguishable, appearing as the curve with the lightest color, which approaches $V_k(\phi)=(1-\phi^2)^2/2$.}
    \label{fig:potphi4}
\end{figure}
The potential \eqref{potphi4} is shown in Fig.~\ref{fig:potphi4}. As $a$ gets larger, the right minimum of the potential is increasingly shifted up, approaching the minimum $\phi=1$ of the potential \eqref{phi4modelkink}. The potential \eqref{potphi4} supports a lump solution that departs from and returns to the minimum $\phi=-1$ after passing through the turning point $\phi= 2\tanh(a)-1$, given by 
$\phi_l(x) = \tanh(x+a)-\tanh(x-a)-1$; this lump can be written as
\be\label{solphi4}
\phi_l(x) = \frac{2\sinh(2a)}{\cosh(2x)+\cosh(2a)}-1.
\ee
Asymptotically, we have $\phi_l(x)=-1+4\sinh(2a)e^{-2|x|} + {\cal O}(e^{-4|x|})$, which is a consequence of the exponential tail of the kink \eqref{phi4modelkink}. Its associated energy density is obtained from \eqref{rhomais}, which reads
\be\label{rhophi4}
\rho_l(x) = \frac{16\sinh^2(2a)\sinh^2(2x)}{(\cosh(2x)+\cosh(2a))^4}.
\ee
By integrating the above expression, we get the energy
\be
E_l=\frac83 - 8\, \csch^2(2a) (2a\coth(2a)-1).
\ee
This energy is monotonically increasing with $a$, approaching the value $E=8/3$ as $a$ increases.

\begin{figure}[t!]
    \centering
    \includegraphics[width=0.5\linewidth]{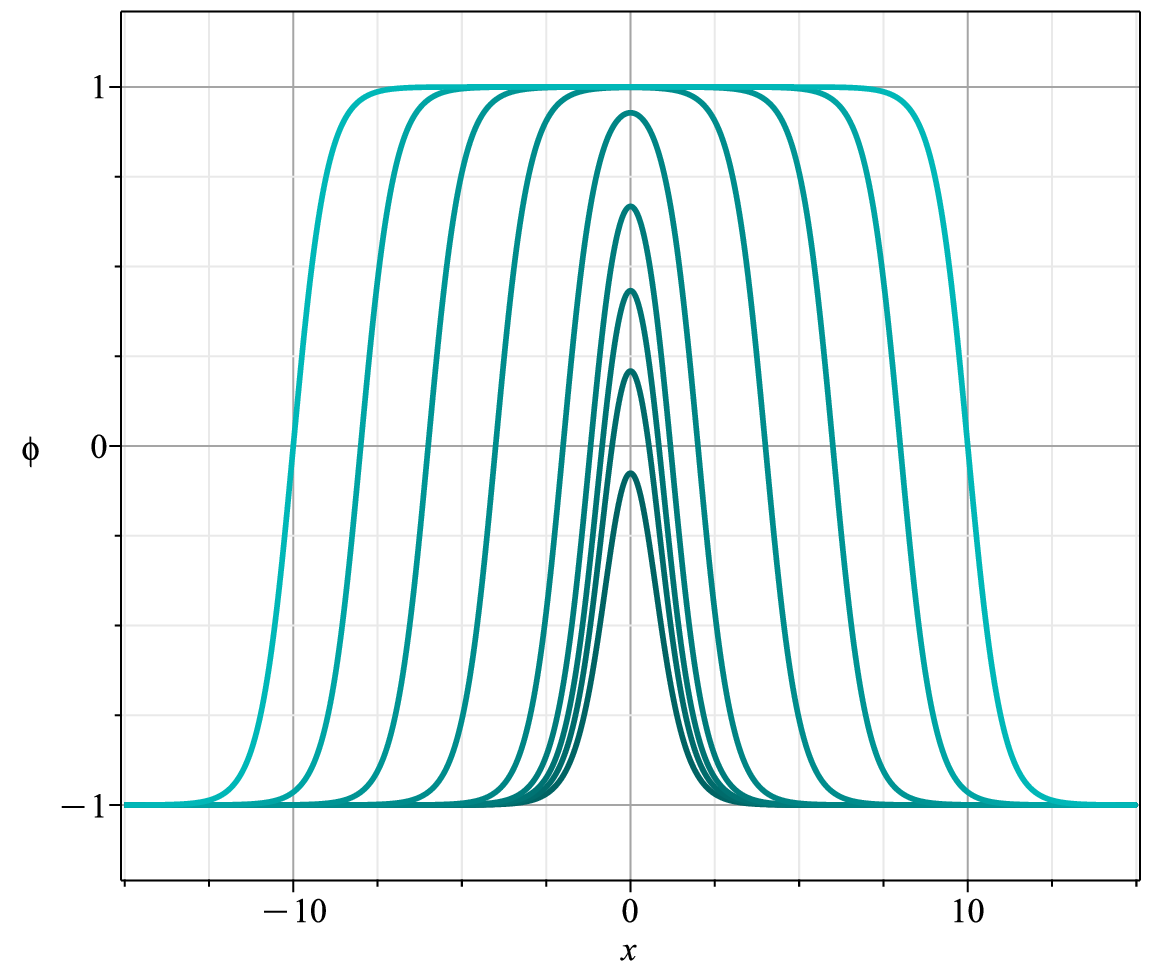}\includegraphics[width=0.5\linewidth]{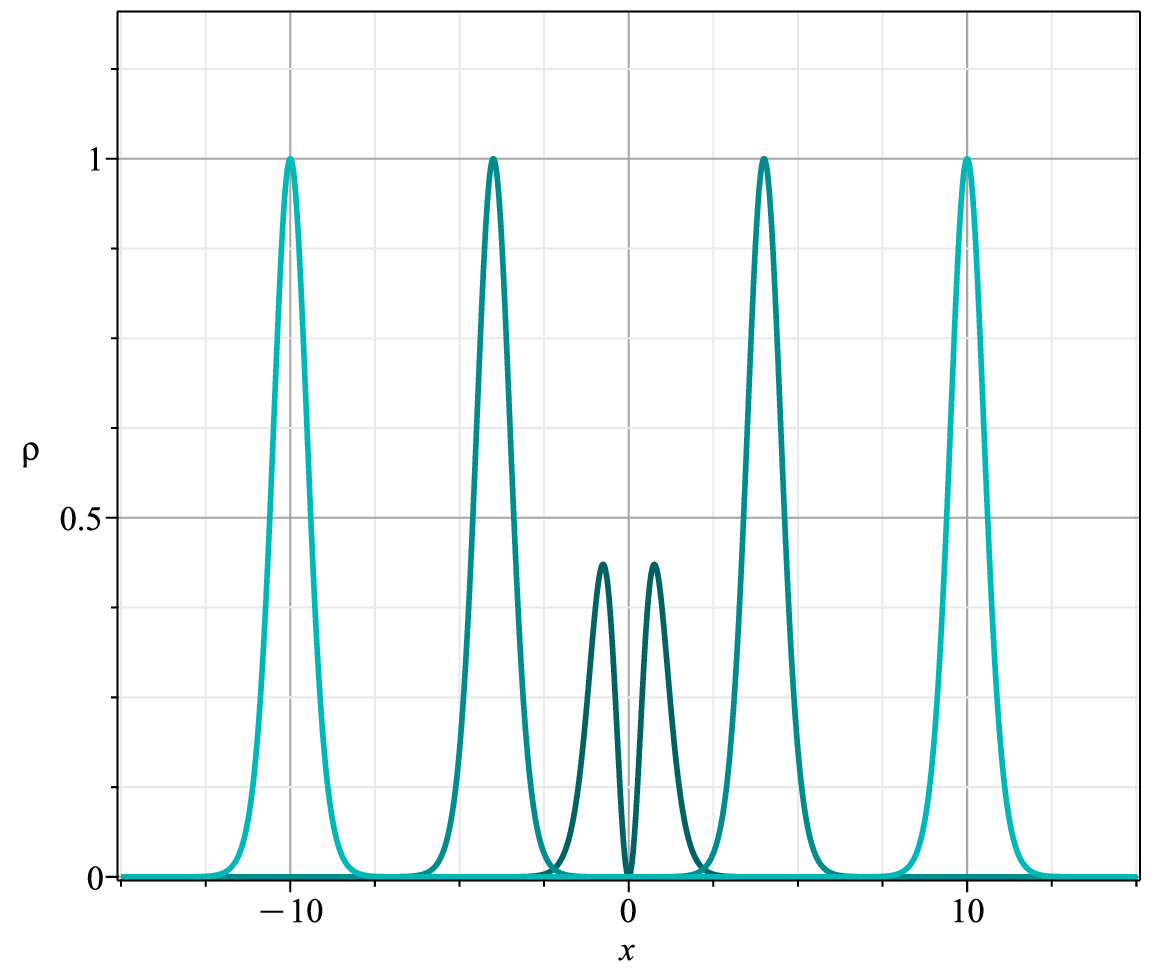}
    \caption{The solution \eqref{solphi4} (left) for $a=0.5,0.7,0.9,1.2,2,4,6,8 $ and $10$, and the energy density \eqref{rhophi4} (right) for $a=0.5, 4$ and $10$.}
    \label{figphi4}
\end{figure}

In Fig.~\ref{figphi4}, we display the solution \eqref{solphi4} and the energy density \eqref{rhophi4}. The effect of increasing $a$ in the solution confirms the emergence of a plateau around the origin, as discussed before, which is a consequence of the separation between the kink and antikink used to obtain the lump. Notice that, as $a$ gets larger, the plateau also gets wider, so the kink/antikink pair becomes widely separated, approximately attaining their individual boundary values. This is evidence that, for very large values of $a$, the kink and antikink work as if they were interacting only via the tails. This behavior can also be seen in the right panel of Fig.~\ref{figphi4}, where the energy density \eqref{rhophi4} shows that, for large $a$, the solution \eqref{solphi4} can indeed be understood as two substructures widely separated.

\subsection{Lumps from sine-Gordon model}
We can also consider the sine-Gordon model in our procedure, with
\be\label{sinegordon}
V_k(\phi)=\frac12\cos^2(\phi),\qquad \phi_k(x)=\arcsin(\tanh x).
\ee
This potential has the set of minima $\phi=(2n-1)\pi/2$, with $n\in\mathbb{Z}$. The above solution was chosen for simplicity; it exists in the central interval $[-\pi/2,\pi/2]$. The energy density of the above kink is $\rho_k(x) = \sech^2(x)$, so the energy is $E_k=2$.

\begin{figure}
    \centering
    \includegraphics[width=0.75\linewidth]{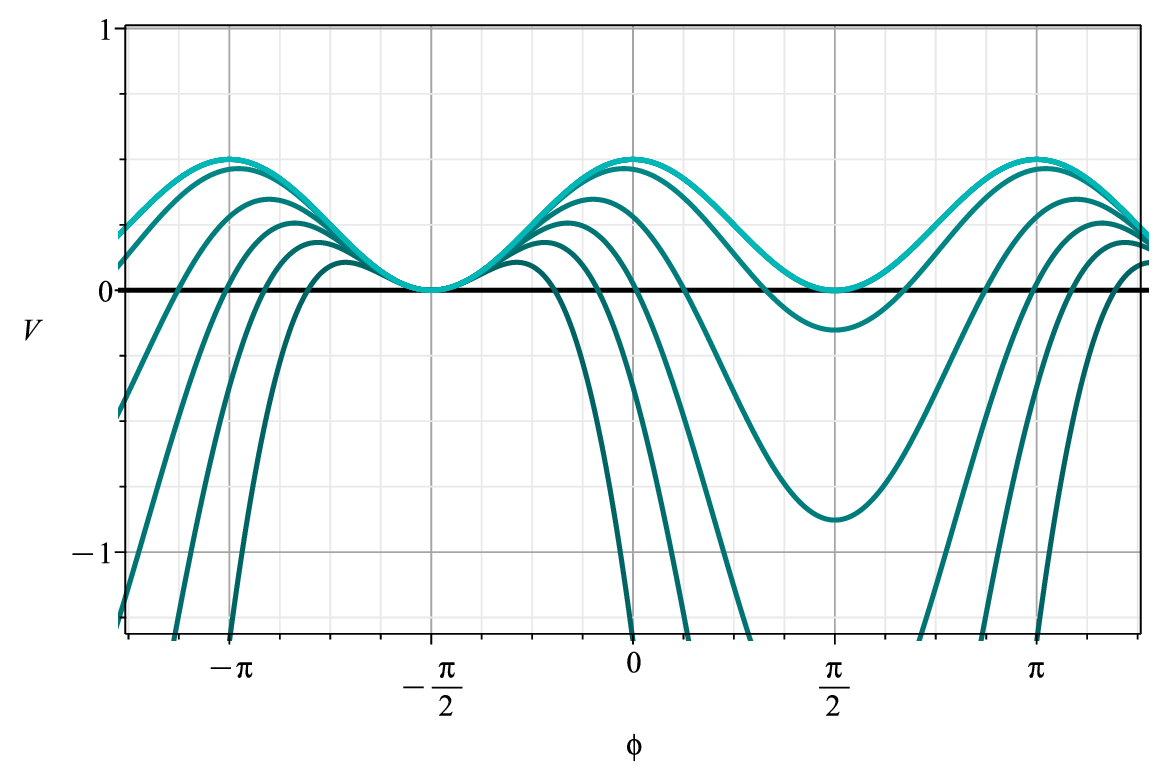}
    \caption{The potential \eqref{potlsg} depicted for $a=0.5,0.7,0.9,1.2,2,4,6,8$ and $10$. The shades of green follow the previous figures. For $a\geq4$, the lines become indistinguishable.}
    \label{fig:potlsg}
\end{figure}

To obtain the lump solution, we use Eq.~\eqref{solmais}, which leads to $\phi_l(x) = \arcsin(\tanh(x+a))-\arcsin(\tanh(x-a))-\pi/2$; it can be written in the form
\be\label{lumpsine}
\phi_l(x)=2\arctan\!\left(\frac{\sinh(a)}{\cosh (x)}\right)-\frac{\pi}{2}.
\ee
Far away from the origin, we get the asymptotics $\phi(x)=-\pi/2+4\sinh(a)e^{-|x|} + {\cal O}(e^{-3|x|})$. We can use the above solution in Eq.~\eqref{fo} to obtain the $2\pi$-periodic potential
\be\label{potlsg}
V_l(\phi)=\frac12\cos^2(\phi)-\frac12\,\csch^2(a)\,(1+\sin (\phi))^2.
\ee
This potential supports two sets of minima, given by $\phi_{\pm} = \pm\pi/2 + 2\pi n$, $n\in\mathbb{Z}$, such that $V(\phi_-)=0$ and $V(\phi_+)=-2\,\csch^2(a)$. As expected, we get $V(\phi_+)\to0$ for $a\to\infty$. The non-minimum zero that is a point of return of the lump solution \eqref{lumpsine} is $\phi_0=2\arctan\!\left(\sinh(a)\right)-\pi/2$. In Fig.~\ref{fig:potlsg}, we display the above potential for several values of $a$.

\begin{figure}[t!]
    \centering
    \includegraphics[width=0.5\linewidth]{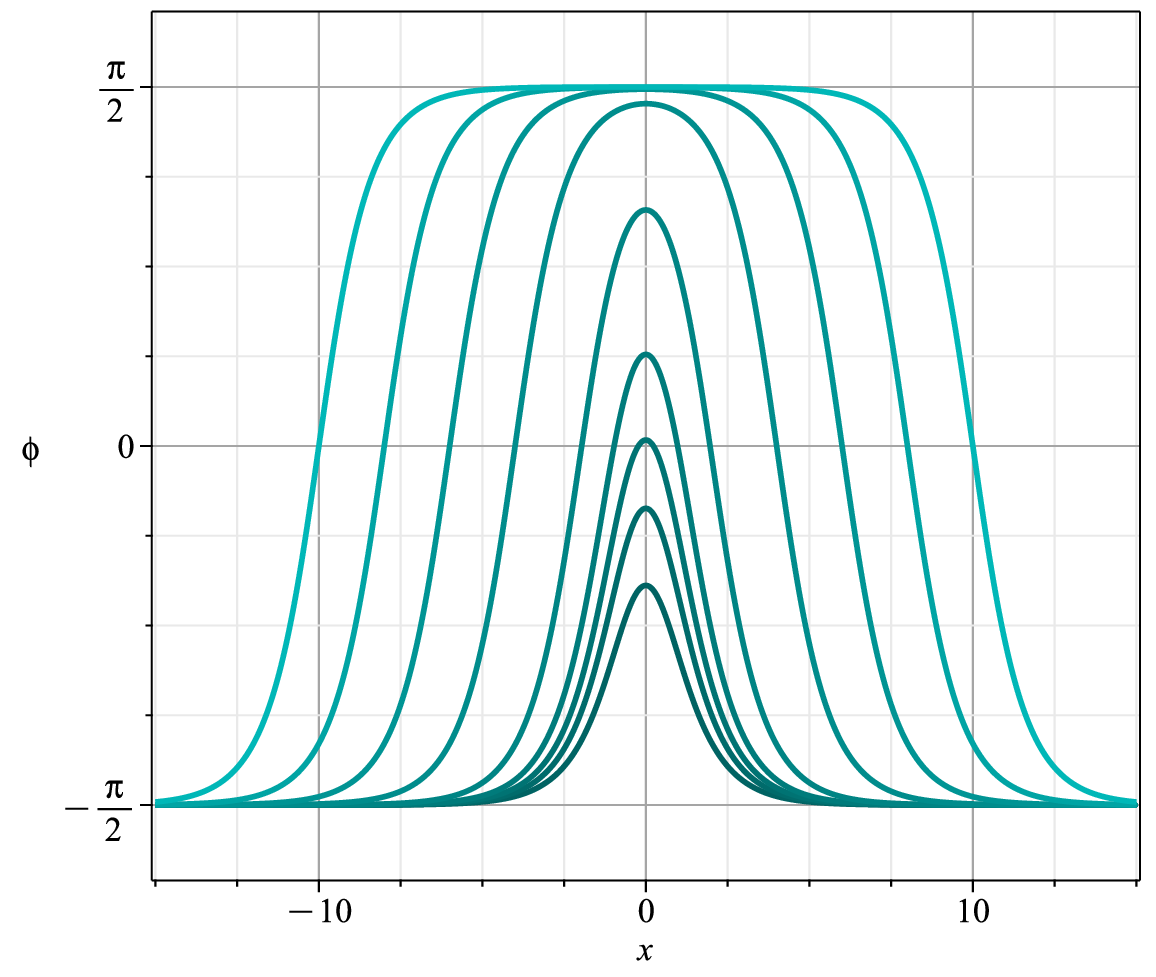}\includegraphics[width=0.5\linewidth]{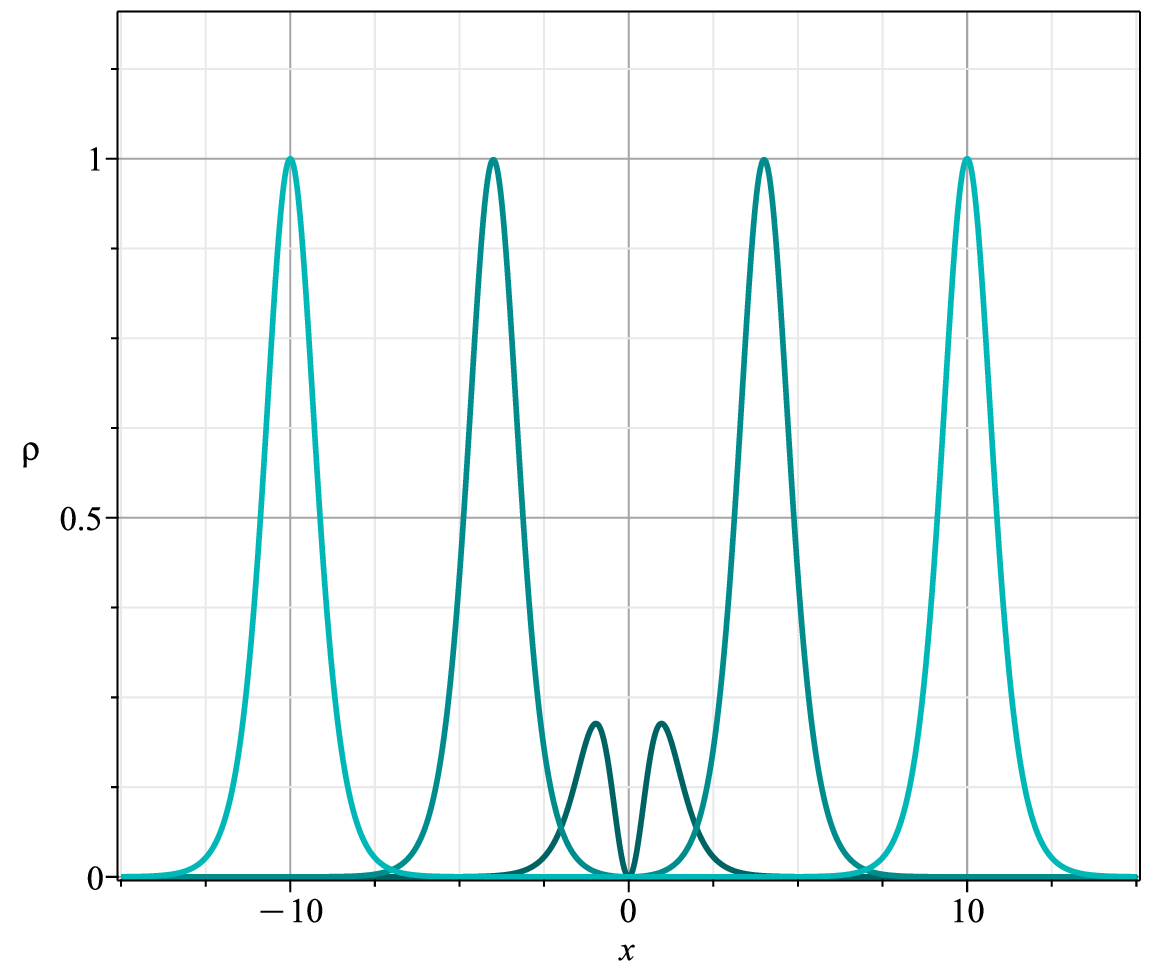}
    \caption{The solution \eqref{lumpsine} (left) for $a=0.5,0.7,0.9,1.2,2,4,6,8 $ and $10$, and the energy density \eqref{rholsg} (right) for $a=0.5, 4$ and $10$. The colors follow the previous figures.}
    \label{figsolsg}
\end{figure}

The energy density \eqref{rho} of the lump solution \eqref{lumpsine} is
\be\label{rholsg}
\rho_l(x)=
\frac{4\,\sinh^2(a)\,\sinh^2(x)}
{\bigl(\cosh^2(x)+\sinh^2(a)\bigr)^2}.
\ee
By integrating it, we get the energy
\be
E_l(a)
=
4\left(1-\frac{2a}{\sinh(2a)}\right).
\ee
As expected, the limit $a\to0$ leads to null energy because it is simply the vacuum $\phi=-\pi/2$ of the sine-Gordon model and $a\to\infty$ leads to $E=4$, which represents the sum of the energies of two infinitely separated kinks. In Fig.~\ref{figsolsg}, we show the behavior of the lump \eqref{lumpsine} and its energy density \eqref{rholsg} for several values of $a$.

\subsection{Lumps from double sine-Gordon model}
The third model under investigation is the double sine-Gordon, described by
\be
V_k(\phi)=\frac2{1+r}\,(r+\cos(\phi))^2,
\ee
with $r\in(0,1)$. For $r=0$, one recovers the sine-Gordon model. The above potential is $2\pi$-periodic, engendering minima at $\phi_\pm=(2n+1)\pi\pm\arccos(r)$. There are two types of maxima, which we call tall and short. For the tall ones, we get $\phi_{\rm tall} = 2n\pi$, with $V(\phi_{\rm tall}) = 2(1+r)$. The short ones appear at $\phi_{\rm short} = (2n+1)\pi$, with $V(\phi_{\rm short}) = 2(1-r)^2/(1+r)$. This feature gives rise to two distinct sets of topological sectors. For simplicity, we only deal with the ones nearer to the origin, which we call $I_{\rm tall} = [-\pi+\arccos(r),\pi-\arccos(r)]$ and $I_{\rm short} = [\pi-\arccos(r),\pi+\arccos(r)]$, respectively corresponding to the kink that passes through the tall and short maxima.

For these sectors, the kink solutions are
\bes
\bal\label{phidsgtall}
&\phi_k^{\rm tall}(x)=2\arctan\!\left(\sqrt{\frac{1+r}{1-r}}\,\tanh\!\left(\sqrt{1-r}\,x\right)\right),\\ \label{phidsgshort}
&\phi_k^{\rm short}(x)=\pi + 2\arctan\!\left(\sqrt{\frac{1-r}{1+r}}\,\tanh\!\left(\sqrt{1-r}\,x\right)\right).
\eal
\ees
Their energies are
\bes
\bal\label{esgk1}
& E^{\rm tall}_{k}=4\sqrt{1-r}+\frac{4r}{\sqrt{1+r}}\left(\pi-\arccos r\right),\\
\label{esgk2}
& E^{\rm short}_{k}=4\sqrt{1-r}-\frac{4r}{\sqrt{1+r}}\arccos r.
\eal
\ees

\begin{figure}[t!]
    \centering
    \includegraphics[width=0.5\linewidth]{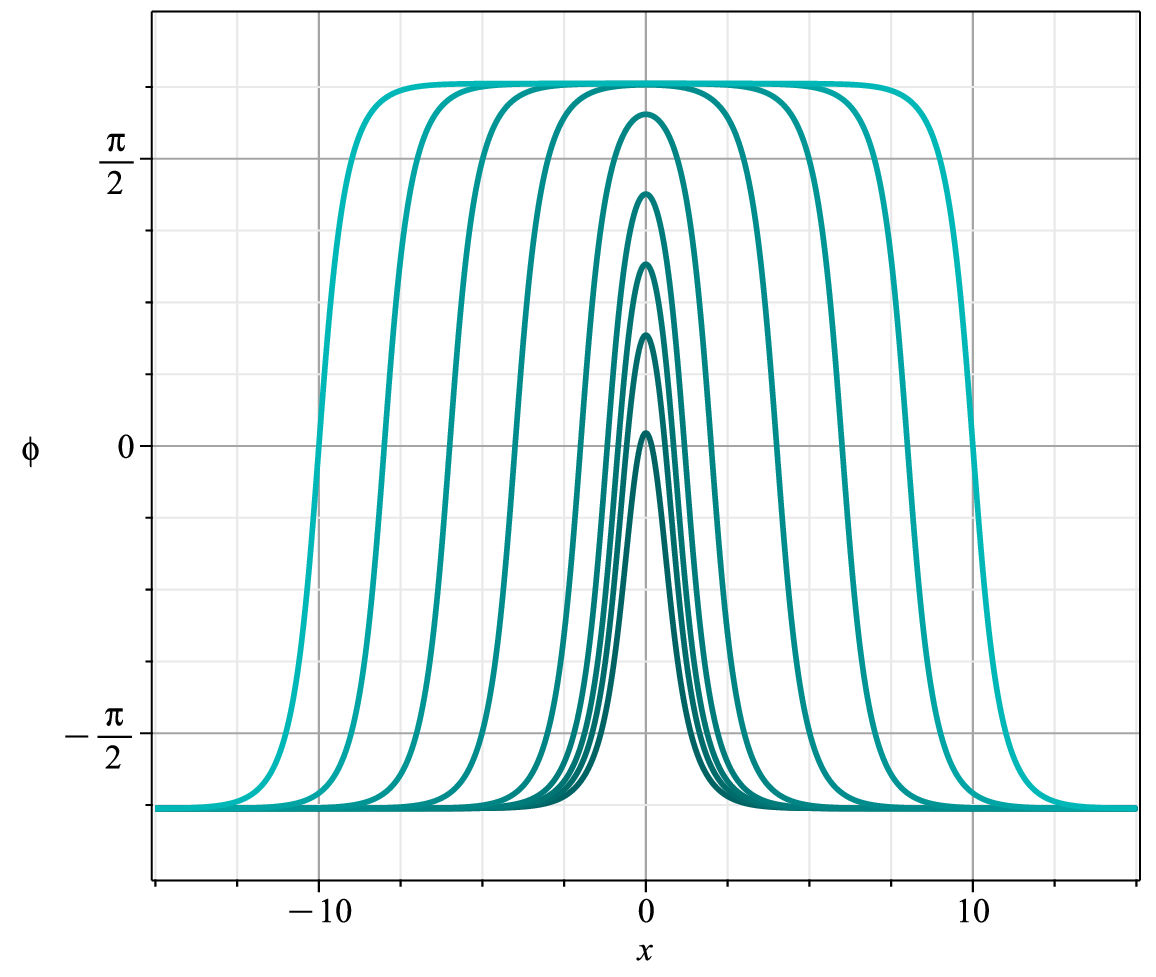}\includegraphics[width=0.5\linewidth]{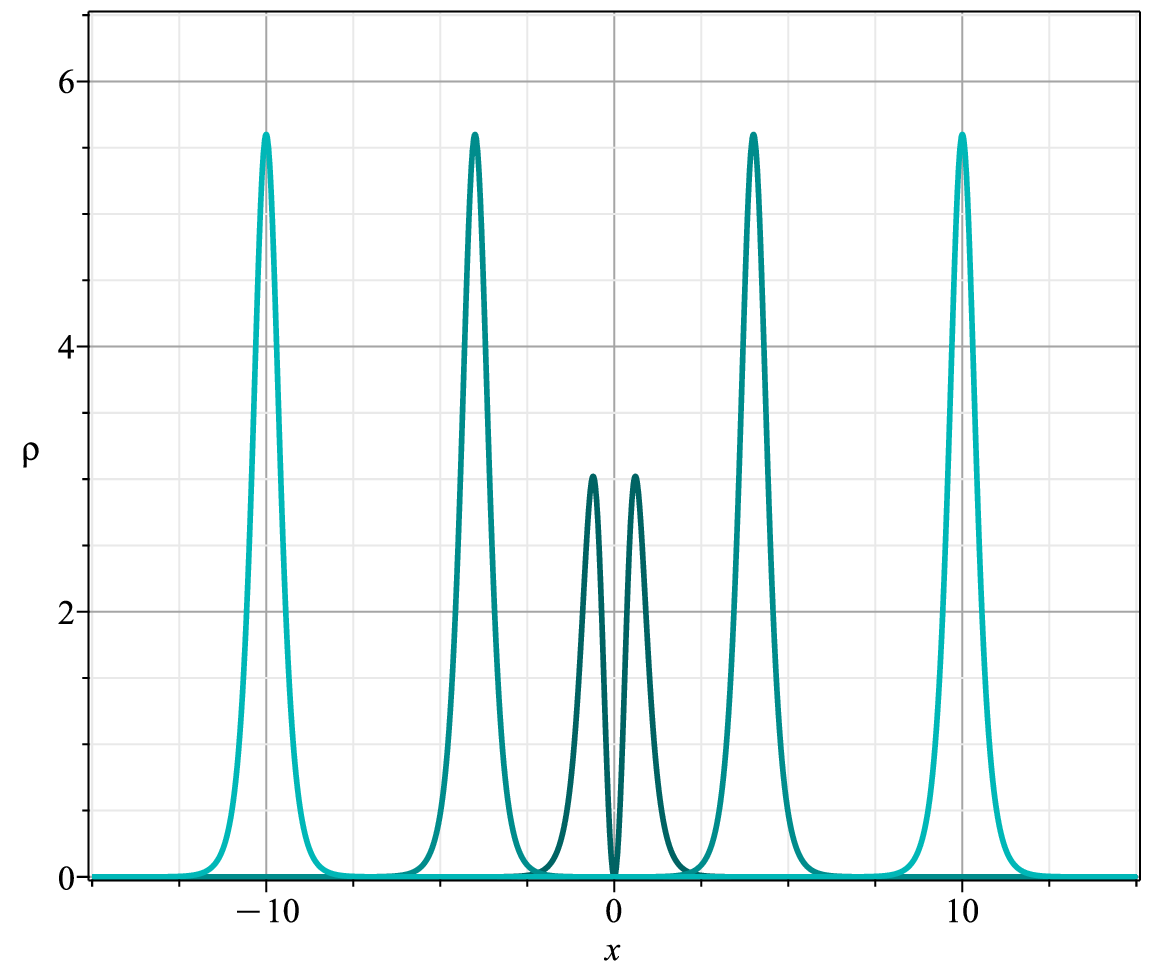}
    \includegraphics[width=0.5\linewidth]{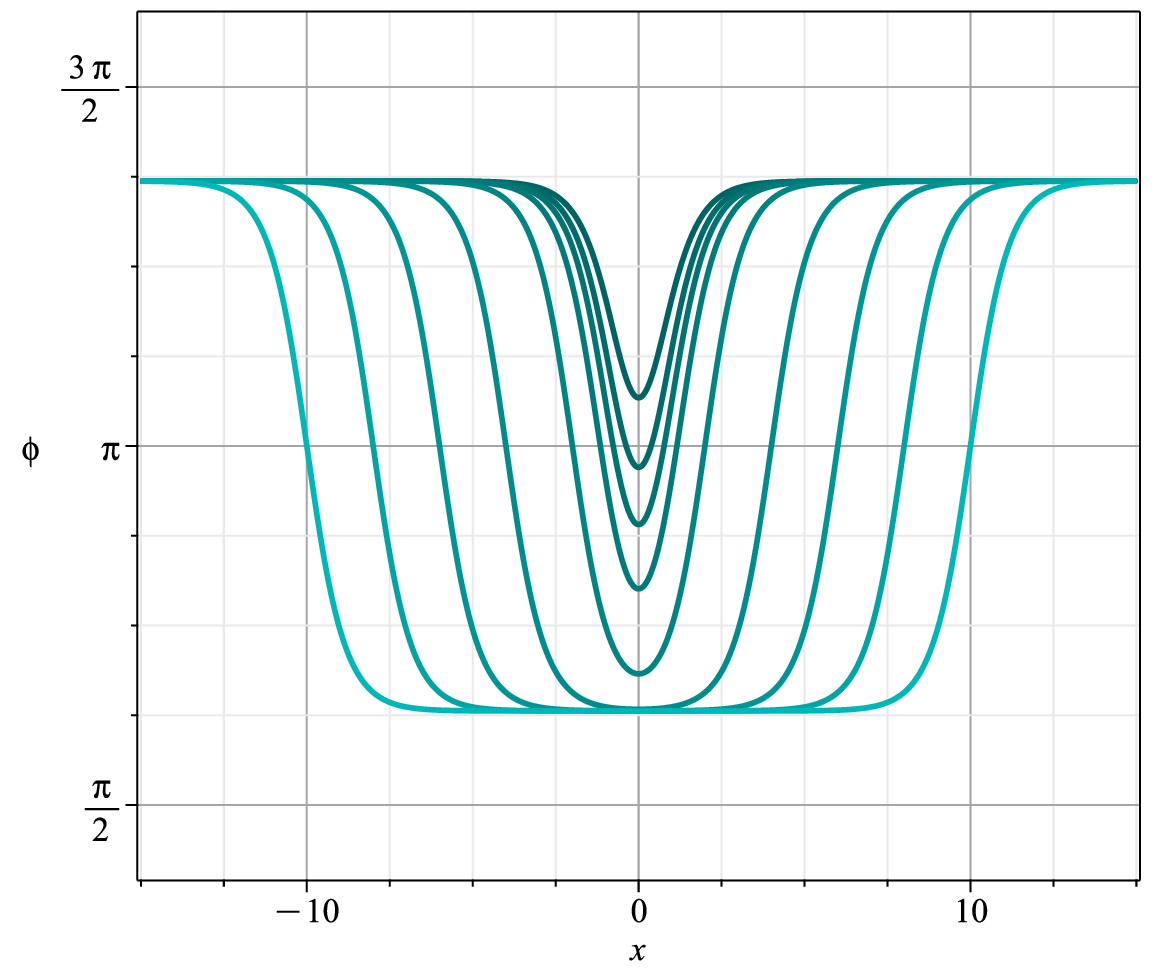}\includegraphics[width=0.5\linewidth]{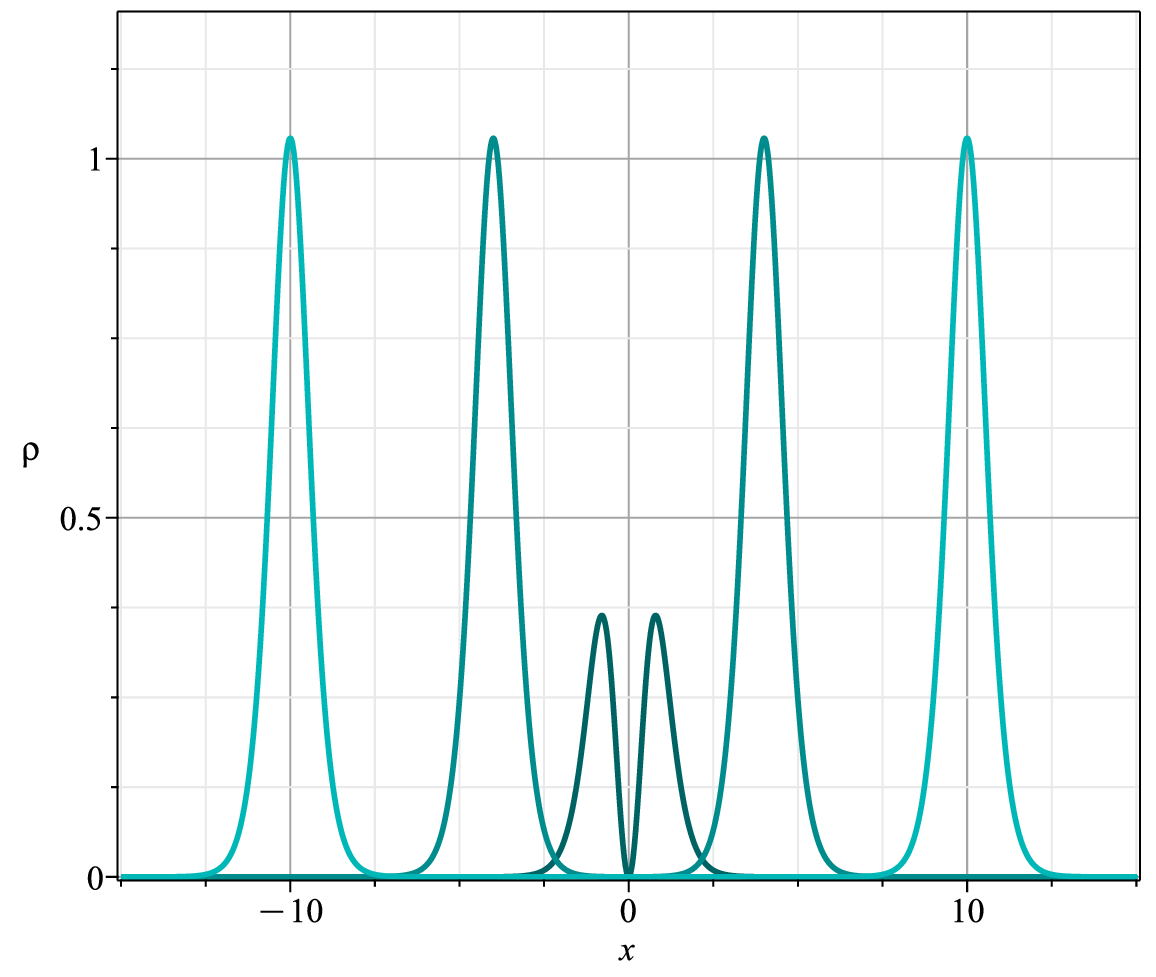}
    \caption{The solutions in Eqs.~\eqref{lumpdsg1} (top left) and \eqref{lumpdsg2} (bottom left) for $a=0.5,0.7,0.9,1.2,2,4,6,8 $ and $10$, and the energy density \eqref{rholdsg1} (top right) and \eqref{rholdsg2} (bottom right) for $a=0.5, 4$ and $10$. We have used $r=0.4$ in all the panels.}
    \label{figsoldsg}
\end{figure}

\begin{figure}
    \centering
    \includegraphics[width=0.75\linewidth]{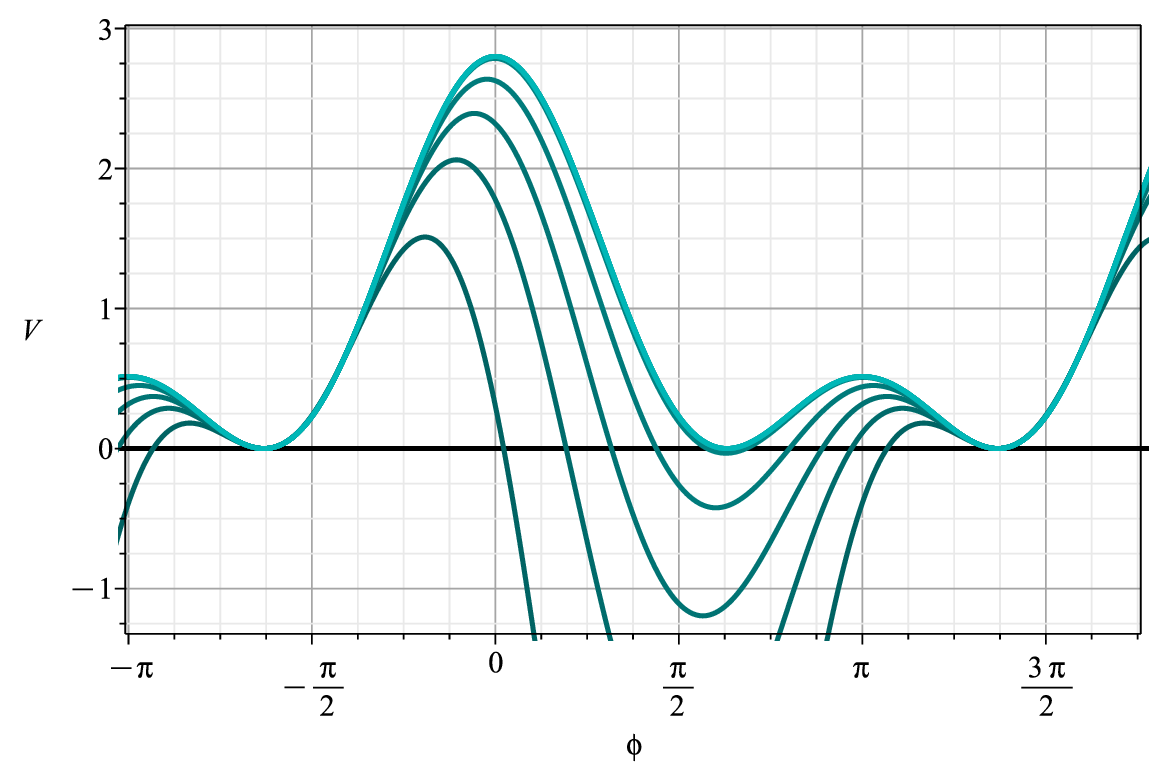}
    \caption{The potential \eqref{potldsg1} with $r=0.4$, depicted for $a=0.5,0.7,0.9,1.2,2,4,6,8$ and $10$. The shades of green follow the previous figures. For $a\geq4$, the lines become indistinguishable.}
    \label{fig:potdsg1}
\end{figure}

We then use our prescription and consider the kink-antikink pair \eqref{solmais}, to get the lump solution. First, we do it with the tall kink \eqref{phidsgtall}, which leads us to
\be\label{lumpdsg1}
\phi^{\rm tall}_{l}(x)=\begin{cases}
\displaystyle
2\arctan\!\left(\frac{s}{Q(x)}\right)-\pi+\arccos r,
& Q(x)>0,
\\[4mm]
\displaystyle
\arccos r,
& Q(x)=0,
\\[4mm]
\displaystyle
2\arctan\!\left(\frac{s}{Q(x)}\right)+\pi+\arccos r,
& Q(x)<0.
\end{cases}
\ee
In this expression, $Q(x)=\cosh(2\beta x) -rc$, with the parameters $\beta=\sqrt{1-r}$, $\gamma=\sqrt{1+r}$, $c=\cosh(2a\beta)$ and $s=\beta\gamma\,\sinh(2a\beta)$. The asymptotic behavior of the above lump is $\phi^{\rm tall}_l(x)=-\pi+\arccos(r)+4\beta\gamma\sinh(2\beta a)e^{-2\beta\,|x|}+{\cal O}(e^{-4\beta\,|x|})$. The energy density is
\be\label{rholdsg1}
\rho^{\rm tall}_{l} = \frac{16\beta^4\gamma^2\,\sinh^2(2\beta x)\sinh^2(2\beta a)}{\big(\cosh(2\beta(x+a))-r\big)^2\big(\cosh(2\beta(x-a))-r\big)^2},
\ee
which can be integrated to lead us to
\be
E^{\rm tall}_{l}=8\beta+\frac{8r(c^2-1)}{\gamma(c^2-r^2)}\bigl(\pi-\arccos r\bigr)-\frac{16a\beta^5\gamma^3\,c}{s(c^2-r^2)}.
\ee
In the limit $a\to\infty$, we get $E^{\rm tall}_{l}=2E^{\rm tall}_{k}$, where $E^{\rm tall}_{k}$ is given by Eq.~\eqref{esgk1}. The solution \eqref{lumpdsg1} and its energy density \eqref{rholdsg1} can be seen in Fig.~\ref{figsoldsg}.
The lump \eqref{lumpdsg1} is a solution of the equation of motion associated with the $2\pi$-periodic potential
\be\label{potldsg1} 
V_{l}(\phi)=\frac{8\beta^{2}\Phi^{2}}{(1+\Phi^{2})^{2}}
\left(1+\frac{1+rc}{s}\,\Phi\right)
\left(1-\frac{1-rc}{s}\,\Phi\right),
\ee
with $\Phi=\cot\left((\arccos(r)-\phi)/2\right)$. The minima that are also zeros are located at $\phi = (2n-1)\pi+ \arccos(r)$. The non-minimum zeros are located at $\phi^{\rm tall}_{0} = \arccos(r) - 2\arctan((1- rc)/s) + 2\pi n$ and $\phi^{\rm short}_{0} = \pi+\arccos(r) - 2\arctan(s/(1+ rc)) + 2\pi n$. This potential is depicted in Fig.~\ref{fig:potdsg1}. The solution in Eq.~\eqref{lumpdsg1} goes asymptotically to the minimum $\phi=-\pi+\arccos(r)$ and has $\phi^{\rm tall}_{0}$ as a point of return; see this behavior in Fig.~\ref{fig:potdsg1}.

We note that there is the possibility of the existence of another lump solution. Indeed, we see in the figure that a solution may depart from and return to the minimum $\phi=\pi+\arccos(r)$ after passing through the point $\phi^{\rm short}_{0}$. It is expected that such a lump solution can be obtained with our prescription \eqref{solmenos} formed by $\phi_{k}^{\rm short}(x)$ in Eq.~\eqref{phidsgshort}, as we want to shift down the left minimum of the potential; see the bottom-left panel of Fig.~\ref{figsoldsg}. This procedure leads to
\be\label{lumpdsg2}
\phi^{\rm short}_{l}(x) = \pi+\arccos(r)-2\arctan\left(\frac{s}{\cosh(2\beta x) +rc}\right).
\ee
It is depicted in the bottom-left panel of Fig.~\ref{figsoldsg}. By constructing the potential from this solution, we obtain a potential with the same form as \eqref{potldsg1}, however changed by $\phi\to 2\pi-\phi$. The energy density of the solution \eqref{lumpdsg2} has the form
\be\label{rholdsg2}  
\rho^{\rm short}_{l} = \frac{16\beta^4\gamma^2\,\sinh^2(2\beta x)\sinh^2(2\beta a)}{\big(\cosh(2\beta(x+a))+r\big)^2\big(\cosh(2\beta(x-a))+r\big)^2},
\ee
which leads to the energy
\be
E^{\rm short}_{l}(a)=8\beta-\frac{8r(c^2-1)}{\gamma(c^2-r^2)}\,\arccos(r)-\frac{16a\beta^5\gamma^3\,c}{s(c^2-r^2)}.
\ee
As before, we also get $E^{\rm short}_{l} = 2E^{\rm short}_{k}$ for $a\to\infty$, with $E^{\rm short}_{k}$ given by Eq.~\eqref{esgk2}. The energy density above can be seen in the bottom-right panel of Fig.~\ref{figsoldsg}.

\subsection{Lumps from a long-range sine-Gordon model}
We consider the starting model as
\be\label{kink2}
V_k(\phi)=\frac12\cos^4(\phi),\quad \phi_k(x) = \arctan(x).
\ee
This model (or at least similar versions of it) was previously investigated in Refs.~\cite{long1,long2,long3}. Similarly to the sine-Gordon model, this potential engenders minima at $\phi=(2n-1)\pi/2$, $n\in\mathbb{Z}$. The above solution $\phi_k(x)$ was chosen for convenience; it exists in the topological sector $[-\pi/2,\pi/2]$. The energy density is $\rho_k(x)=(1+x^2)^{-2}$, which leads to $E_k=\pi/2$.

Considering the kink in \eqref{kink2}, we use \eqref{solmais} to write $\phi(x)=\arctan(x+a)-\arctan(x-a)-\pi/2$, or
\be\label{lumparctan}
\phi_l(x)=
\begin{cases}
\arctan\!\left(\dfrac{2a}{Q(x)}\right)-\dfrac{\pi}{2},
& Q(x)>0,\\[2ex]
0,
& Q(x)=0,\\[2ex]
\arctan\!\left(\dfrac{2a}{Q(x)}\right)+\dfrac{\pi}{2},
& Q(x)<0.
\end{cases}
\ee
The denominator that appears above is $Q(x)=1+x^2-a^2$. This lump departs from $\phi=-\pi/2$, touches the point $\phi=\phi_0\equiv2\arctan(a)-\pi/2$ and returns to $\phi=-\pi/2$. The asymptotic behavior of this solution is given by
$\phi_l(x)\approx -\pi/2 +2a/x^2+\mathcal{O}\left(1/x^4\right)$,
in which $x\gg a$. We can use \eqref{lumparctan} in Eq.~\eqref{fo} to construct the potential
\be\label{vsglong}
V_l(\phi) = \frac{1}{2}\cos^4(\phi)\left(1-\frac{1}{a^2}-\frac{2}{a}\tan(\phi)\right).
\ee
For general $a$, it supports inflection points with null derivative, which are all zeros, at $\phi=(n-1/2)\,\pi$, for $n\in\mathbb{Z}$. Other points of interest are located at $\phi=\arctan((a^2-1\pm\sqrt{a^4+a^2+1})/(3a)) +n\pi$, where the positive sign represents the minima and the negative sign stands for the maxima. The non-extremal zeros appear at $\phi=\phi_0 +n\pi$. As $a$ gets larger, both $\phi_0$ and the minima approach the inflection points and the maximum goes to $n\pi$. In the limit $a\to\infty$, we get $V(\phi)=\cos^4(\phi)/2$, as expected, so the aforementioned minima and $\phi_0$ merge into the inflection points, which become minima at $\phi=(n-1/2)\pi$.
\begin{figure}[t!]
    \centering
    \includegraphics[width=0.75\linewidth]{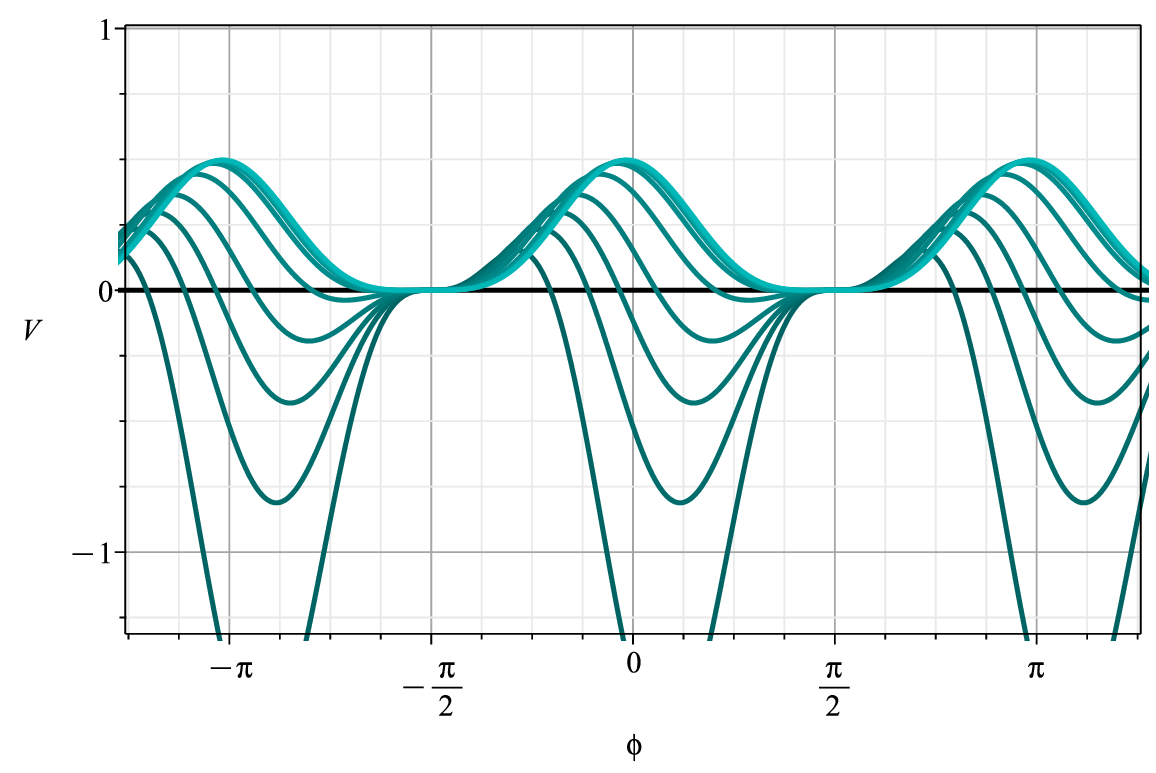}
    \caption{The potential \eqref{vsglong} depicted for $a=0.5,0.7,0.9,1.2,2,4,6,8$ and $10$. The shades of green follow the previous figures. For $a\geq4$, the lines become indistinguishable.}
    \label{fig:potsglong}
\end{figure}
This potential is displayed in Fig.~\ref{fig:potsglong}. The lump \eqref{lumparctan} goes asymptotically to the inflection points with null derivative.

\begin{figure}[t!]
    \centering
    \includegraphics[width=0.75\linewidth]{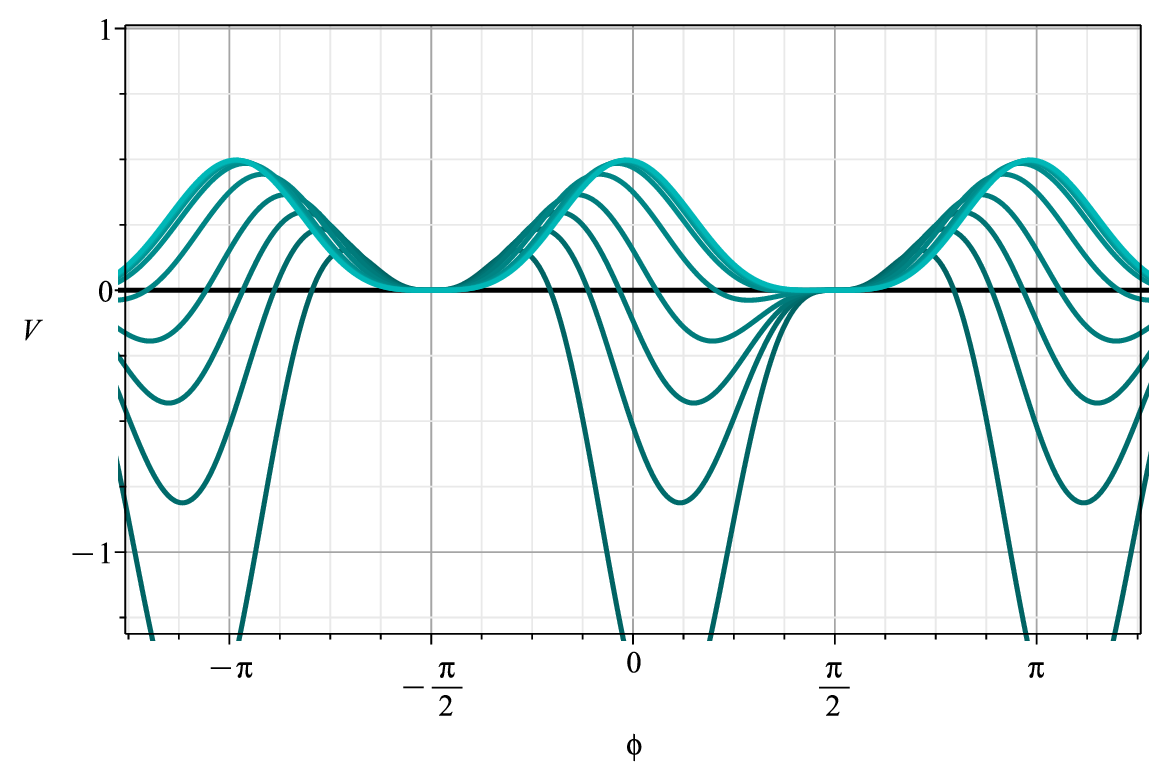}
    \caption{The potential \eqref{vsglongalt} depicted for $a=0.5,0.7,0.9,1.2,2,4,6,8$ and $10$. The shades of green follow the previous figures. For $a\geq4$, the lines become indistinguishable.}
    \label{fig:potsglongalt}
\end{figure}

The absence of a true vacuum can lead to additional instabilities in the model. However, we can circumvent this issue by modifying the potential \eqref{vsglong} to ensure that $\phi = -\pi/2$ is a vacuum. We then write
\be\label{vsglongalt}
\Tilde{V}_l(\phi) =\begin{cases}
	\cfrac{\cos \left(\phi \right)^{4} \left(1-\cfrac{1}{a^{2}}+\cfrac{2 \tan \left(\phi \right)}{a}\right)}{2} & \phi <-\cfrac{\pi}{2},
\\
 \cfrac{\cos \left(\phi \right)^{4} \left(1-\cfrac{1}{a^{2}}-\cfrac{2 \tan \left(\phi \right)}{a}\right)}{2} & \phi\geq-\cfrac{\pi}{2}.
\end{cases}
\ee
Therefore, the lump solution lives in the interval $\phi \in [-\pi/2,\phi_0]$, tending asymptotically to the vacuum $\phi=-\pi/2$ and having $\phi=\phi_0$ as the point of return (see Fig.~\ref{fig:potsglongalt}). One may perform similar modifications to make other inflection points become vacua. 

\begin{figure}[t!]
    \centering
    \includegraphics[width=0.5\linewidth]{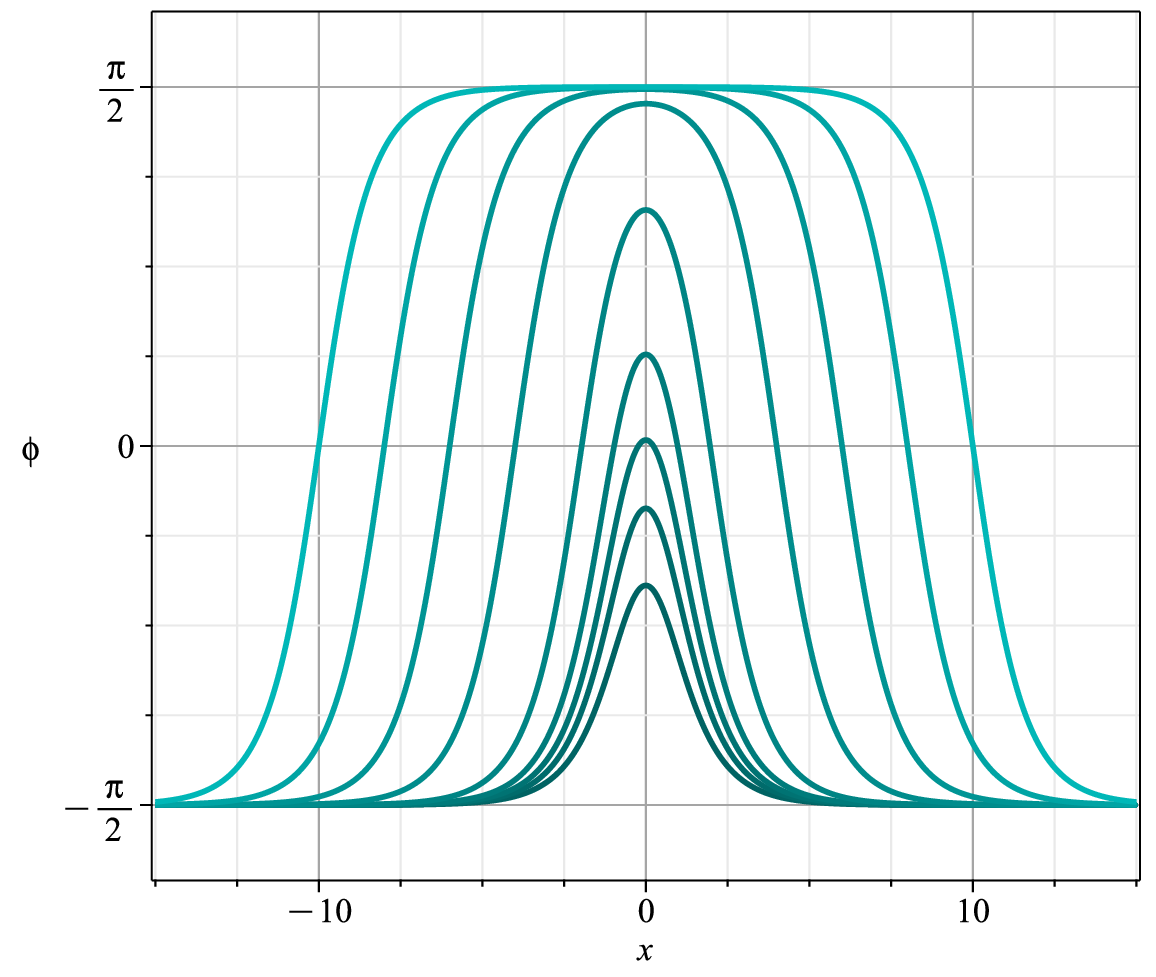}\includegraphics[width=0.5\linewidth]{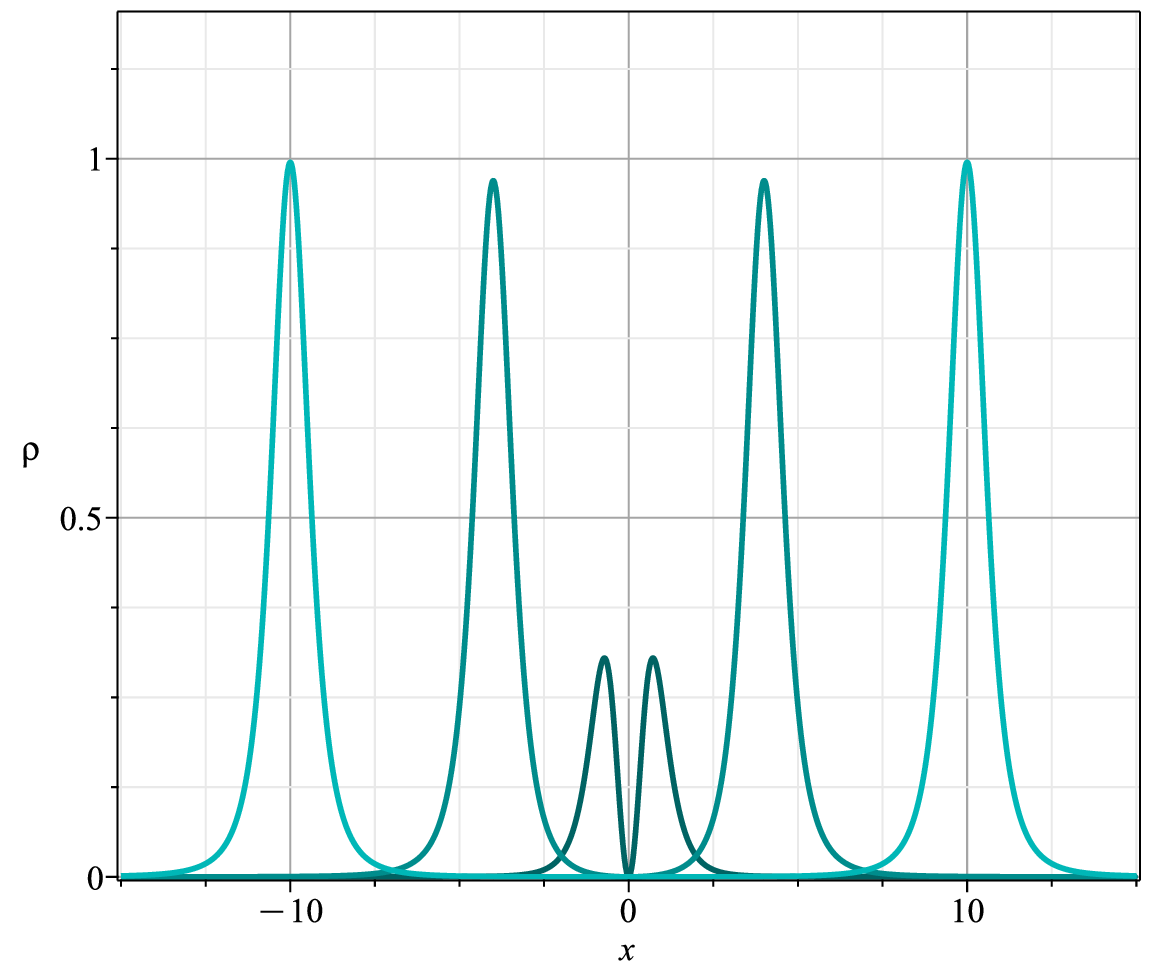}
    \caption{The solution \eqref{lumparctan} (left) for $a=0.5,0.7,0.9,1.2,2,4,6,8 $ and $10$, and the energy density \eqref{rhosglong} (right) for $a=0.5, 4$ and $10$. The shades of green follow the previous figures.}
    \label{figsolsglong}
\end{figure}

By using Eq.~\eqref{rhomais}, we see that the solution \eqref{lumparctan} has energy density
\be\label{rhosglong}
\rho_l(x)=
\frac{16a^2x^2}
{\Bigl[(1+x^2-a^2)^2+4a^2\Bigr]^2},
\ee
whose asymptotic behavior is $\rho_l(x)=16a^2/x^6 + {\cal O}(1/x^8)$. The integral of the above expression over all space leads to the energy
\be
E_l=\frac{\pi a^2}{1+a^2}.
\ee
We get $E_l=0$ for $a\to0$ and $E_l=\pi$ for $a\to\infty$. In Fig.~\ref{figsolsglong}, we show the behavior of the solution \eqref{lumparctan} and its energy density \eqref{rhosglong} for several values of $a$.

\section{Superposition of asymmetric kinks}\label{secasymmetric}
\subsection{A novel model of asymmetric kinks}\label{secasymmetrickinks}
\begin{figure}[t!]
    \centering
    \includegraphics[width=0.75\linewidth]{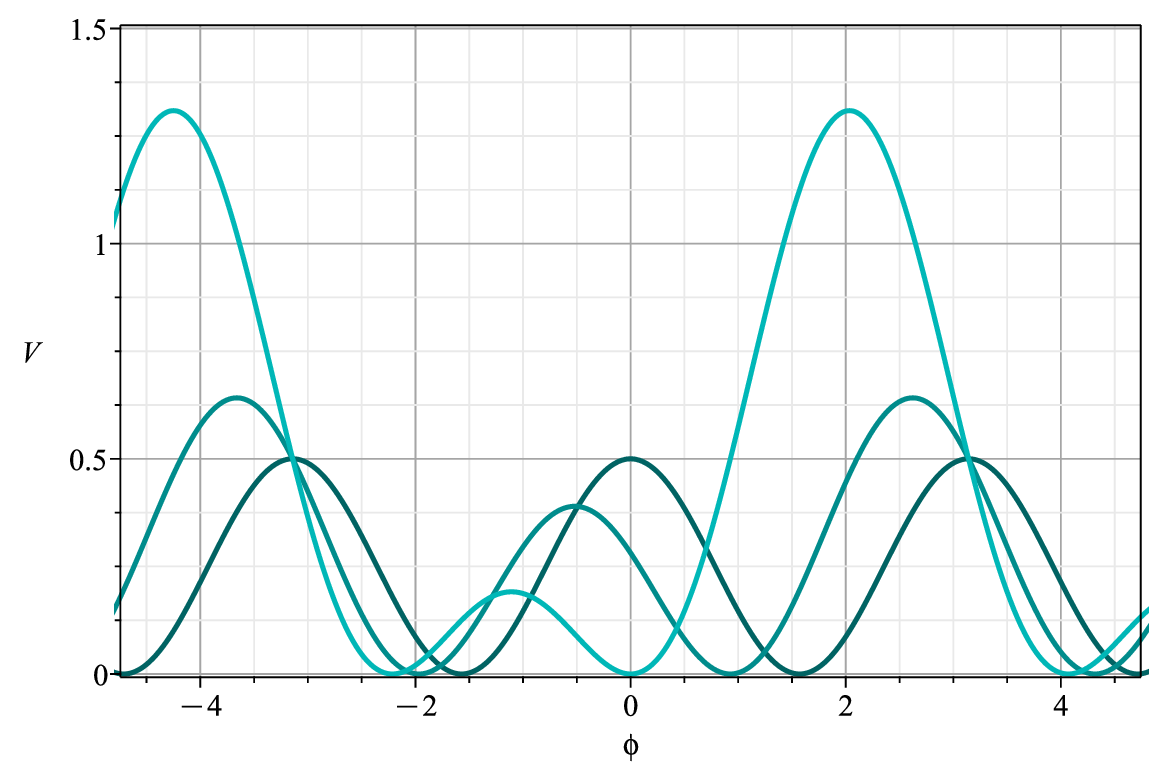}
    \caption{The potential \eqref{potkasy} depicted for $\nu=0,0.5$ and $1$. The shades of green get darker as $\nu$ increases.}
    \label{figpotkasy}
\end{figure}

There are several models in the literature that support asymmetric kinks; see, for instance, Ref. \cite{polynomialinteractions}. However, they do not lead to analytic (and elegant) lump solutions. We then introduce a novel model, described by
\begin{equation}\label{potkasy}
V_k(\phi)=\frac12\left[\cos(\phi)-\nu\sin(\phi)-\nu^2\cos^2\left(\frac\phi2\right)\right]^2,
\end{equation}
where $\nu$ is a positive real parameter responsible for the asymmetry in the kink solutions. Notice that $\nu=0$ recovers the sine-Gordon model \eqref{sinegordon}. Negative values of $\nu$ can be obtained straightforwardly via the change $\phi\to-\phi$ in the potential. In Fig.~\ref{figpotkasy} we display the above potential for some values of $\nu$.
The above expression can be obtained with the deformation procedure \cite{deformation} applied to the sine-Gordon model \eqref{sinegordon} with the deformation function 
\be\label{deformationfunction}
f(\phi)=2\operatorname{arctan}\left(\tan\left(\frac\phi2\right)+\nu\right).
\ee
The potential \eqref{potkasy} has minima at
\be\label{minimaasy}
\phi^{(n)}_\pm=-2\arctan(\nu\mp1)+2\pi n,
\ee
with $n\in\mathbb{Z}$. Two families of maxima with different heights are present. The short ones are located at $\phi_{\rm short} = -\operatorname{atan2}(\nu,1-\nu^2/2)+2\pi n$ and the tall ones at $\phi_{\rm tall} = -\operatorname{atan2}(\nu,1-\nu^2/2)+(2n+1)\pi$, for which we get $V(\phi_{\rm short}) = (1/2)(\sqrt{1+\nu^4/4} - \nu^2/2)^2$ and $V(\phi_{\rm tall}) = (1/2)(\sqrt{1+\nu^4/4} + \nu^2/2)^2$. The notation $\operatorname{atan2}$ stands for the $2$-argument arctangent. These features allow us to see that the potential \eqref{potkasy} supports two types of topological sectors, similarly to the double sine-Gordon model. Notwithstanding that, there is no value for the parameter $\nu$ that makes neighbor minima join into a single point in the potential \eqref{potkasy}, contrary to what occurs in the double sine-Gordon potential for $r\geq1$.

In the topological sector, $\phi\in[\phi^{(n)}_-,\phi^{(n)}_+]$, the above potential supports the following kink and antikink solutions
\bes\label{kinkasy}
\bal
&\phi_k(x)=2\arctan\!\left(\tanh\left(\frac{x}{2}\right)-\nu\right) + 2\pi n,\\
&\phi_{ak}(x)=2\arctan\!\left(-\tanh\left(\frac{x}{2}\right)-\nu\right) + 2\pi n.
\eal
\ees
We can investigate the asymptotic behavior of the above kink solution. Contrary to the previous examples, it engenders an asymmetric profile, with both of its tails being governed by exponentials of same intensity. This is a consequence of the classical masses of the potential \eqref{potkasy}, which are all the same at the minima $(m^{(n)}_\pm=1)$. As we have discussed below Eq.~\eqref{deltasexponential}, the kinks support exponential tails, and here we get $\delta_L(x) = 4e^{x}/d_L + {\cal O}(e^{2x})$ and $\delta_R(x) = -4e^{-x}/d_R+ {\cal O}(e^{-2x})$, where
\be\label{dldr}
d_L=\nu^2+2\nu+2 \quad\text{and}\quad d_R = \nu^2-2\nu+2.
\ee
So, we see that the asymmetry is manifested in the tails via their amplitude, since $d_L\neq d_R$ for $\nu>0$.

Both solutions have the same energy density, given by
\be\label{rhokasy}
\rho_k(x)=
\frac{\operatorname{sech}^4\left(x/2\right)}
{\left[1+\left(\tanh\left(x/2\right)-\nu\right)^2\right]^2}.
\ee
By integrating it, we get
\begin{equation}\label{energykasy}
E_k
=
2-\nu^2\arctan\left(\frac{2}{\nu^2}\right).
\end{equation}
Notice that, for $\nu=0$, Eqs.~\eqref{potkasy}--\eqref{energykasy} recover the corresponding ones for the sine-Gordon model, as expected.

\begin{figure}[t!]
    \centering
    \includegraphics[width=\linewidth]{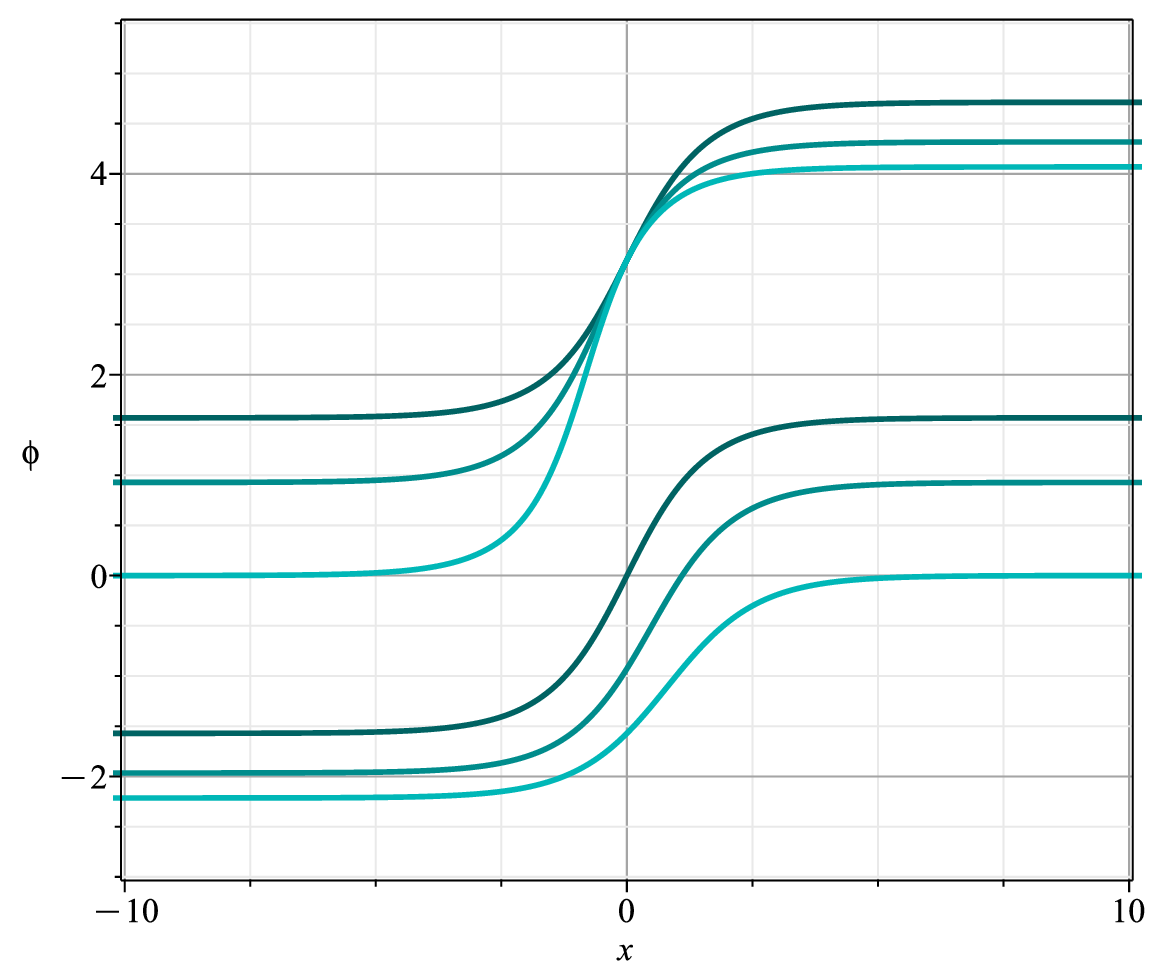}
    \includegraphics[width=0.5\linewidth]{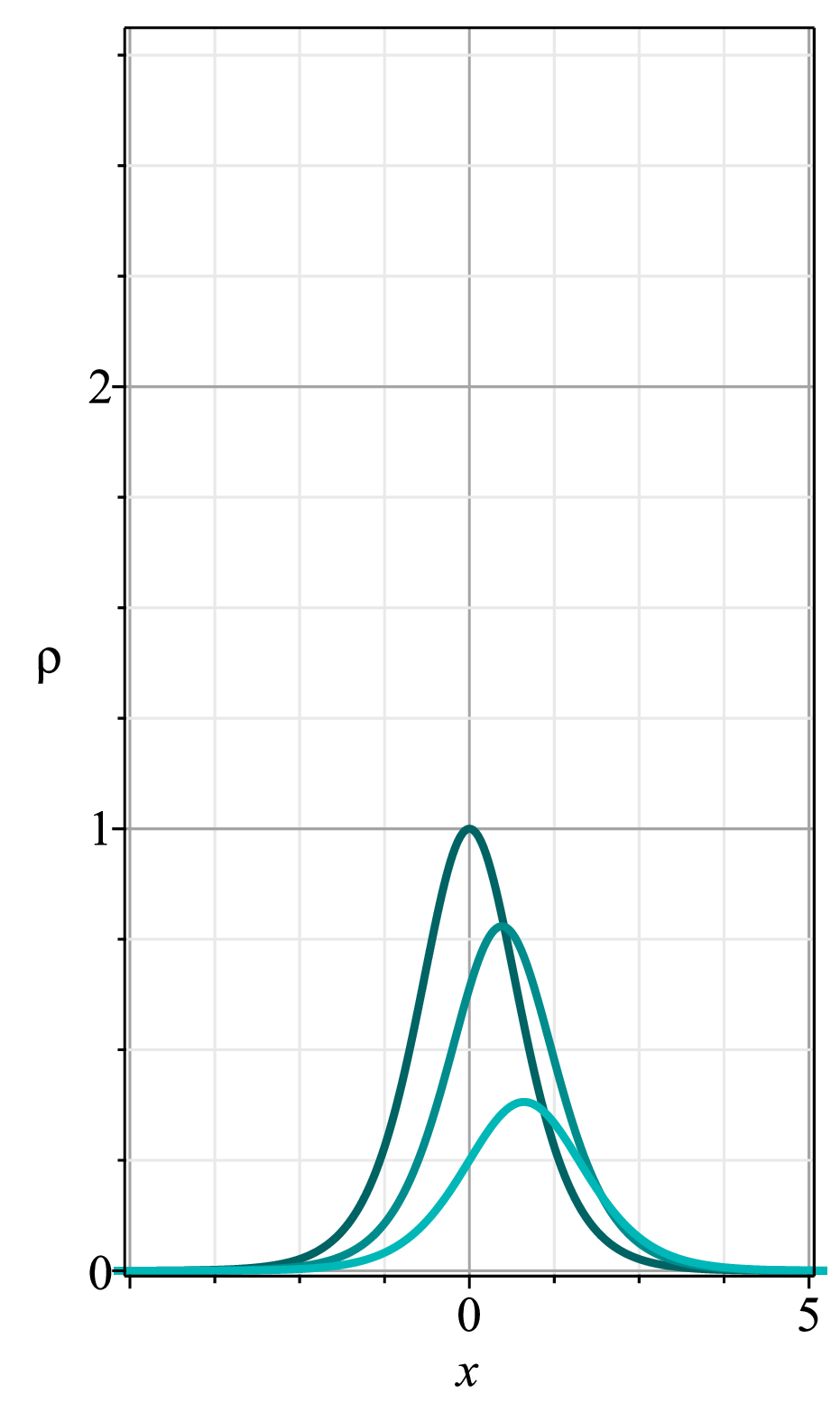}\includegraphics[width=0.5\linewidth]{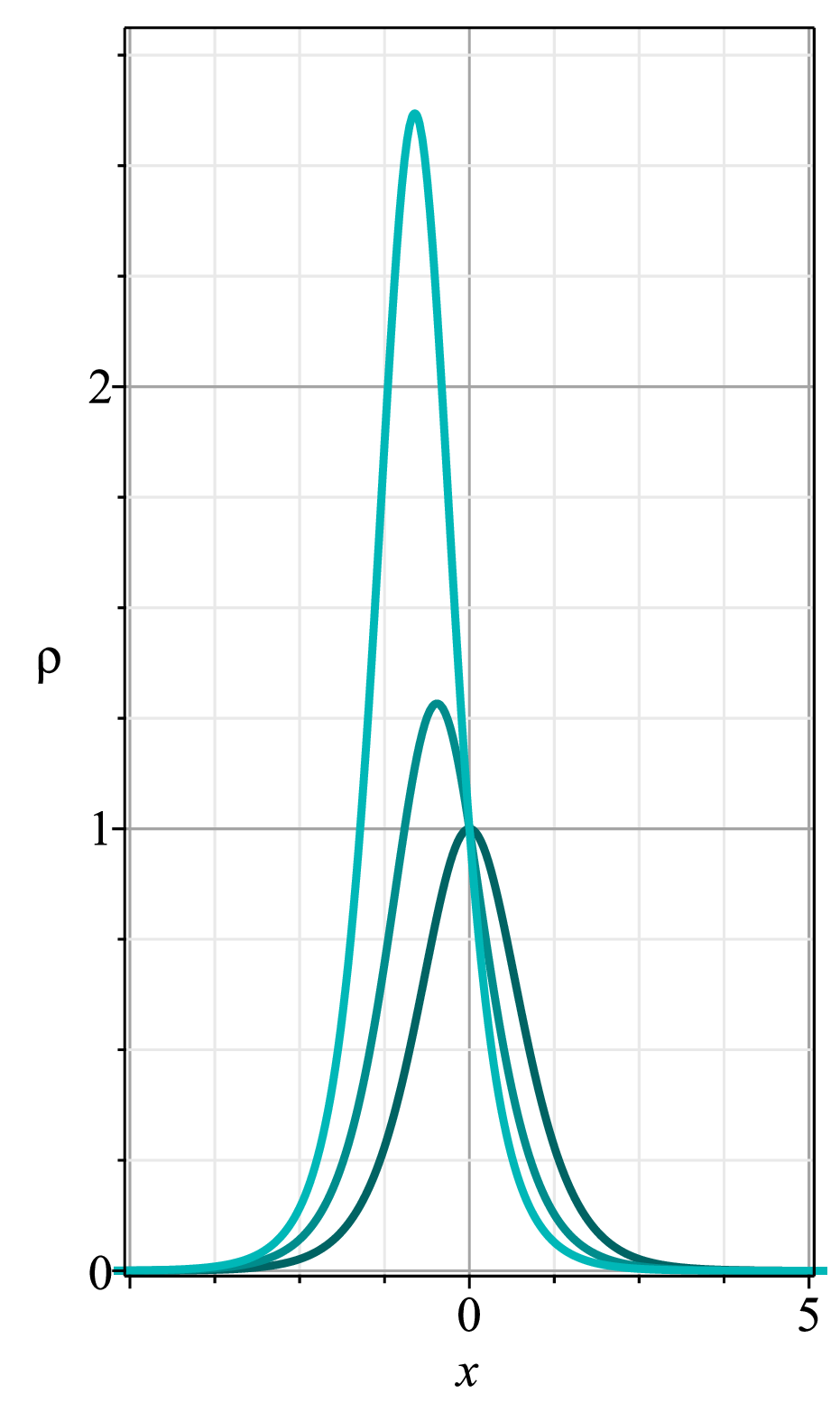}
    \caption{In the top panel, we display the solutions \eqref{kinkasy} (lower curves) and \eqref{soltilde} (upper curves) for $\nu=0,0.5$ and $1$. In the bottom panels, we depict their corresponding energy densities \eqref{rhokasy} (left) and \eqref{rhokasytilde} (right). The colors follow the previous figure.}
    \label{figsolk}
\end{figure}

There is another topological sector of the potential \eqref{potkasy}, $\phi\in[\phi^{(n)}_+,\phi^{(n+1)}_-]$, which supports a different kink, given by
\be\label{soltilde}
\Tilde{\phi}_k(x)=2\pi n
+
\begin{cases}
2\arctan\left(-\coth\dfrac{x}{2}-\nu\right),
& x<0,
\\[0.8em]
\pi,
& x=0,
\\[0.8em]
2\pi
+
2\arctan\left(-\coth\dfrac{x}{2}-\nu\right),
& x>0.
\end{cases}
\ee
The antikink can be obtained via the change $x\to-x$ in the above expression. The associated energy density is
\be\label{rhokasytilde}
\tilde{\rho}_k(x)=
\frac{\operatorname{sech}^4\left(\frac{x}{2}\right)}
{\left[\tanh^2\left(\frac{x}{2}\right)+\left(1+\nu\tanh\left(\frac{x}{2}\right)\right)^2\right]^2}.
\ee
By integrating it, we get the energy
\begin{equation}\label{energykasytilde}
\tilde{E}_k
=
2-\nu^2\arctan\left(\frac{2}{\nu^2}\right) + \pi\nu^2.
\end{equation}
In Fig.~\ref{figsolk}, we display the solutions \eqref{kinkasy} and \eqref{soltilde} and their corresponding energy density \eqref{rhokasy} and \eqref{rhokasytilde} for some values of $\nu$.

\subsection{Lumps}\label{seclumpsasykinks}
The asymmetry of the kink solution requires us to investigate the lumps \eqref{lumpgeral} with care. We then have
\bal\label{philplusasy}
&\phi_l^+(x)=\phi_k(x+a)+\phi_{ak}(x-a)+2\arctan(\nu-1),\\
&\phi_l^-(x)=\phi_{ak}(x+a)+\phi_k(x-a)+2\arctan(\nu+1),
\eal
in which we are considering the kink-antikink pair formed by the solutions \eqref{kinkasy} with $n=0$, for simplicity. For
\be\label{acritasy}
a=a^*\equiv\frac{1}{2}\ln\left(
\frac{\nu^2+2\nu+2}{\nu^2-2\nu+2}
\right),
\ee
the kink and antikink that form the lump \eqref{philplusasy} cancel out, making the solution uniform, so we restrict our analysis to the case $a>a^*$, which is the one of our interest since we are investigating large lumps. These lump configurations are distinct; $\phi_l^+(x)$ returns to the vacuum $\phi^{(0)}_-=-2\arctan(\nu+1)$ and $\phi_l^-(x)$ returns to the vacuum $\phi^{(0)}_+=-2\arctan(\nu-1)$.

These expressions can be summarized in the form
\be\label{lumpasy}
\phi_l^\pm(x)=
2\arctan\!\left(
\frac{P_0^\pm + P_1^\pm \tanh^2(x/2)}
{Q_0^\pm + Q_1^\pm \tanh^2(x/2)}
\right),
\ee
with the coefficients given by
\bes
\bal
& P_0^{\pm}=\pm(1 - 2q - q^2)+ \nu(1 + 2q + q^2)\nonumber\\
&\hspace{0.9cm}\mp \nu^2(1 + 2q) + \nu^3,\\
& P_1^{\pm}=\pm(1 + 2q - q^2)- \nu(1 + 2q + q^2)\nonumber\\
&\hspace{0.9cm}\pm \nu^2 q(2 + q)- \nu^3 q^2,\\
&Q_0^{\pm}=-1 - 2q + q^2\pm 2\nu- \nu^2,\\
&Q_1^{\pm}=-1 + 2q + q^2\mp 2\nu q^2+ \nu^2 q^2,
\eal
\ees
where $q=\tanh({a}/{2})$. The asymptotic behavior of the lump \eqref{lumpasy} can be written in the form
\be
\delta^\pm(x) = 4\left(\frac{e^{\pm a}}{d_L} - \frac{e^{\mp a}}{d_R}\right) e^{-|x|} + {\cal O}\big(e^{-2|x|}\big),
\ee
where $d_L$ and $d_R$ are given by Eq.~\eqref{dldr}. If one tries to make the first contribution zero, the result is \eqref{acritasy}, which makes no sense here as the solution is actually uniform.

\begin{figure}[t!]
    \centering
    \includegraphics[width=0.75\linewidth]{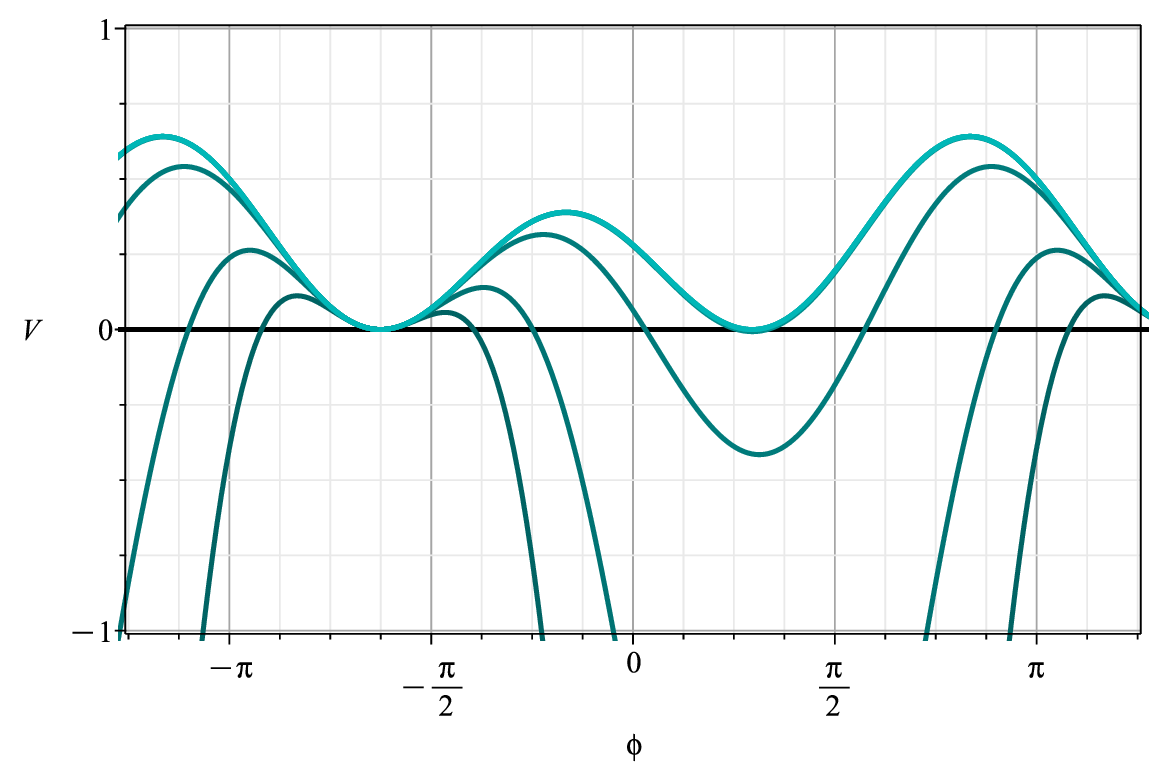}
    \includegraphics[width=0.75\linewidth]{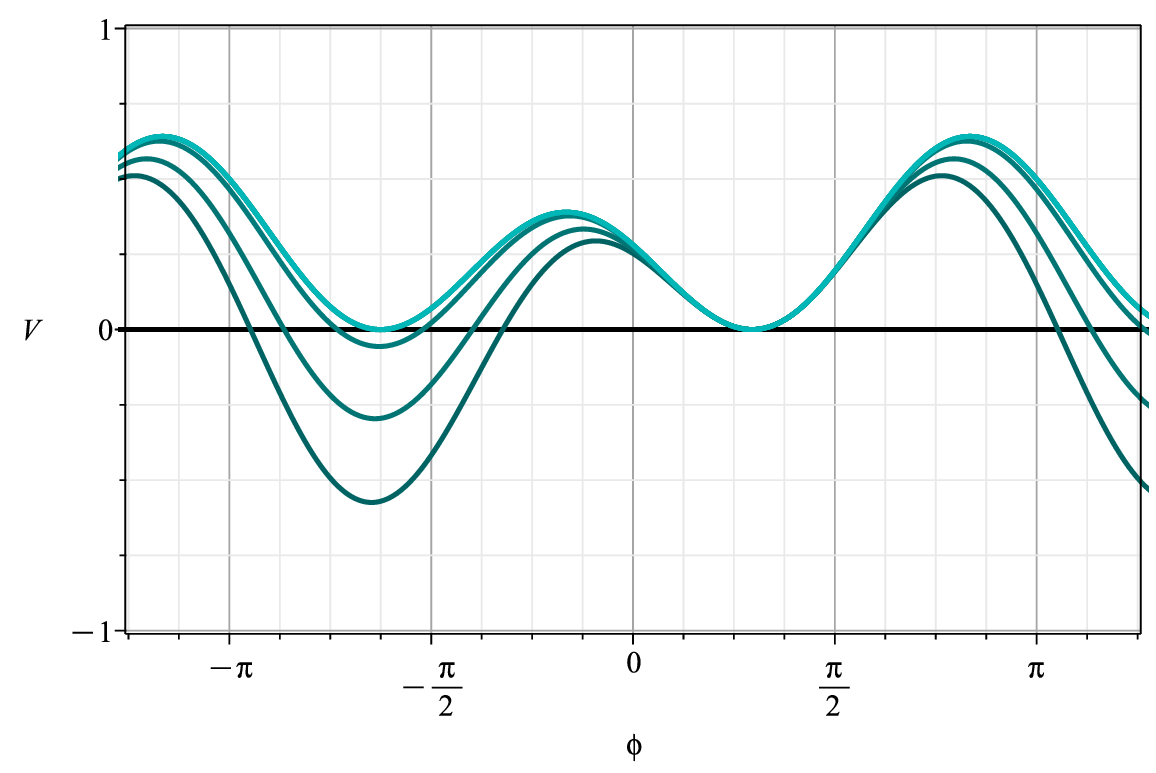}
    \caption{The potentials $V^+_l(\phi)$ (top) and $V^-_l(\phi)$ (bottom) in \eqref{potpmasy} depicted for $\nu=0.5$ and $a=0.9,1.2,2,4,6,8 $ and $10$. The shades of green follow the previous figures. }
    \label{figpotlasy}
\end{figure}

The potential associated with the lump $\phi_l^\pm(x)$ above is
\begin{equation}\label{potpmasy}
V_l^{\pm}(\phi)=\frac{\left(Q_0^{\pm}u - P_0^{\pm}c\right)\left(u + (\nu \pm 1)c\right)^2\left(P_1^{\pm}c - Q_1^{\pm}u\right)}{2M_{\pm}^2},
\end{equation}
with $u=\sin({\phi}/{2})$, $c=\cos({\phi}/{2})$ and
$M_{\pm} = \mp2q +\nu (1+q^2) \mp \nu^2q$. In Fig.~\ref{figpotlasy}, we display these potentials for $\nu=0.5$ and several values of $a$.
The energy density of the lump solution $\phi_l^\pm(x)$ is
\be\label{rholasypm}
\rho_l^\pm(x)=\frac{4K_\pm^2\,T^2S^4}{\left[\left(P_0^\pm+P_1^\pm T^2\right)^2+\left(Q_0^\pm+Q_1^\pm T^2\right)^2\right]^2},
\ee
with $K_\pm=P^\pm_1Q^\pm_0-P^\pm_0Q^\pm_1$, $T=\tanh(x/2)$ and $S=\operatorname{sech}(x/2)$. If we integrate the above equation, we get a cumbersome expression for the energy. So, we omit it here.

\begin{figure}[t!]
    \centering
    \includegraphics[width=0.5\linewidth]{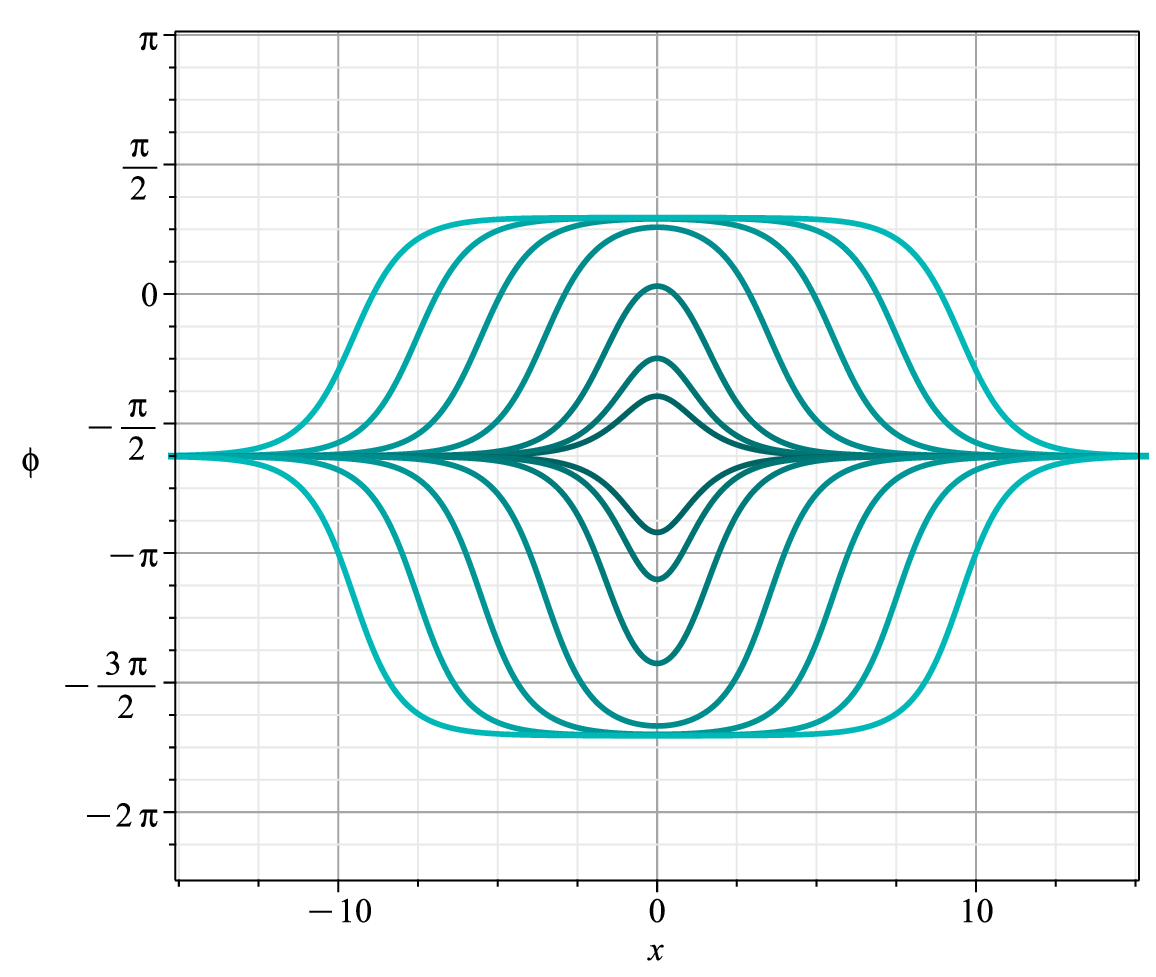}\includegraphics[width=0.5\linewidth]{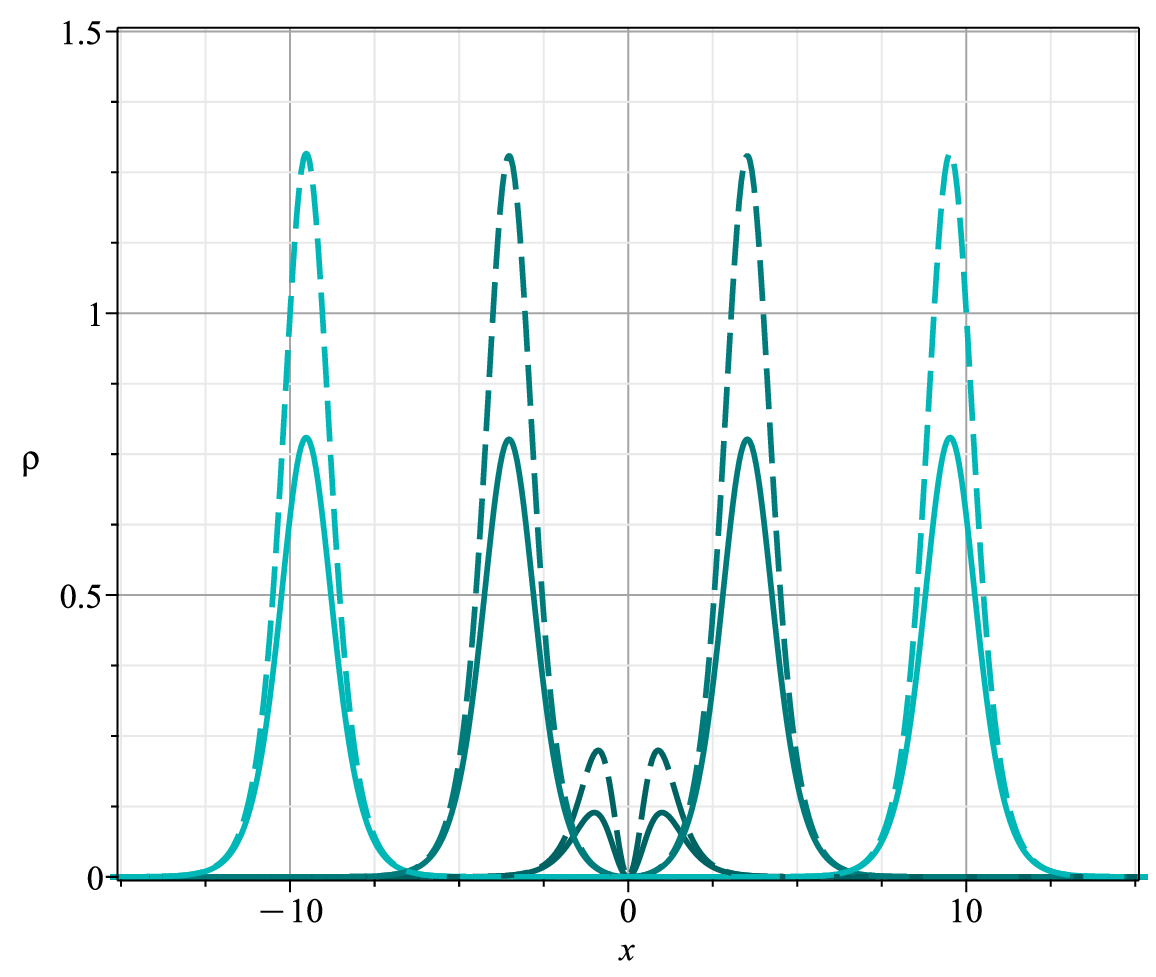}
    \caption{In the left panel, we display the solutions $\phi^+(x)$ in \eqref{lumpasy} and $\Tilde{\phi}^-(x)$ in Eq.~\eqref{lumptildem} for $\nu=0.5$ and $a=0.9,1.2,2,4,6,8 $ and $10$. In the right panel, we depict their corresponding energy density $\rho^+(x)$ in \eqref{rholasypm} (solid lines) and  $\Tilde{\rho}^-(x)$ in \eqref{rhotildeasypm} (dashed lines), for $a=0.9,4$ and $10$. The colors follow the previous figures.}
    \label{figsollasy1}
\end{figure}

\begin{figure}[t!]
    \centering
    \includegraphics[width=0.5\linewidth]{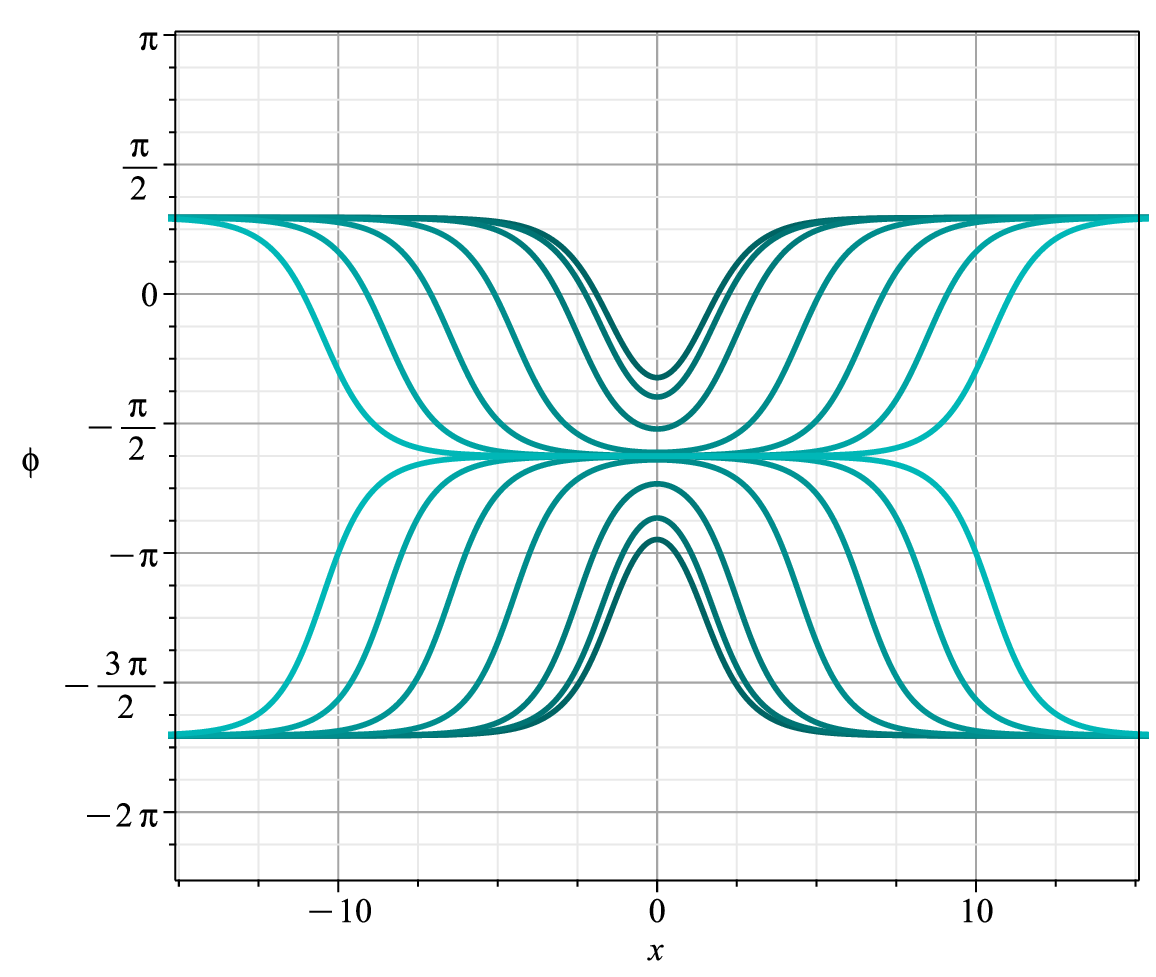}\includegraphics[width=0.5\linewidth]{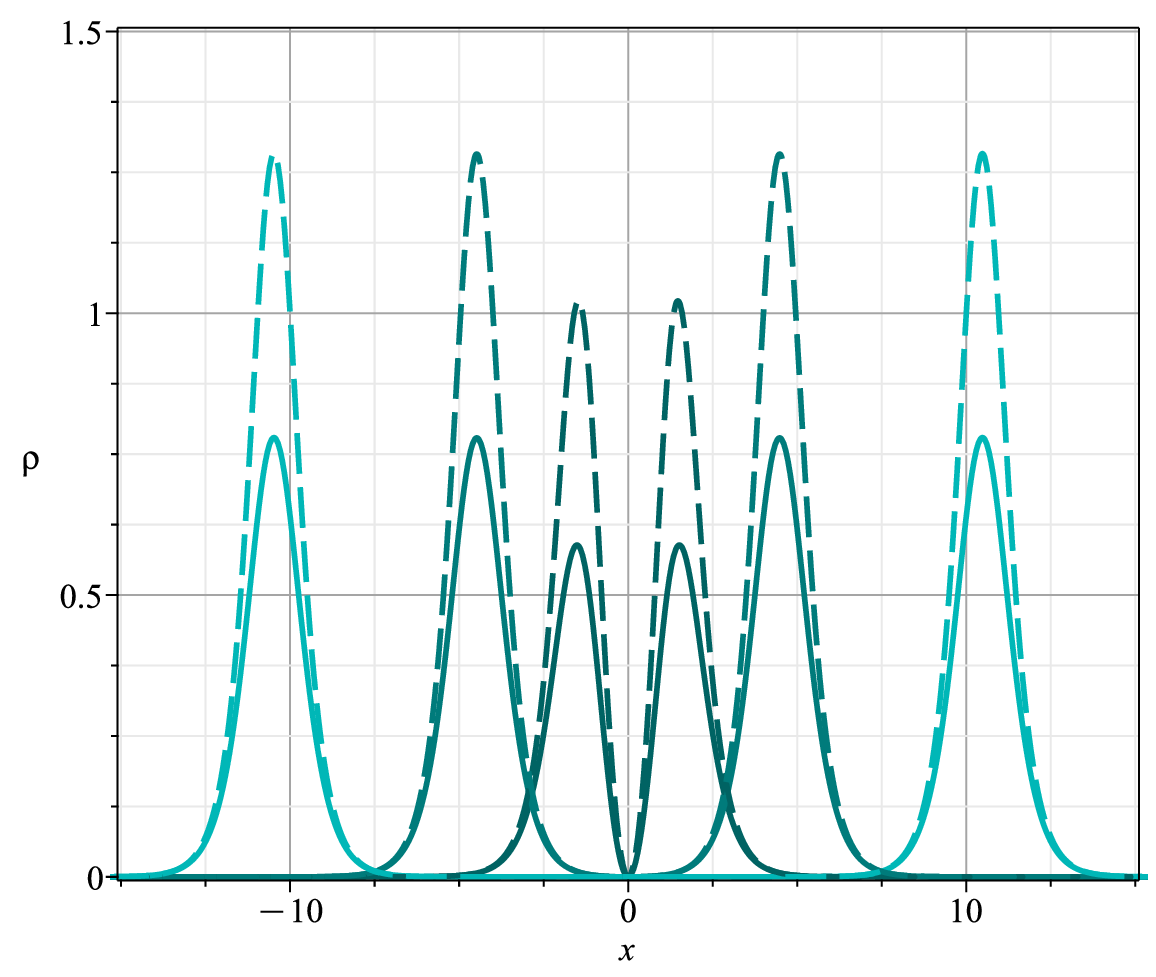}
    \caption{In the left panel, we display the solutions $\phi^-(x)$ in \eqref{lumpasy} and $\Tilde{\phi}^+(x)$ in Eq.~\eqref{lumptildep} for $\nu=0.5$ and $a=0.9,1.2,2,4,6,8 $ and $10$. In the right panel, we depict their corresponding energy density $\rho^-(x)$ in \eqref{rholasypm} (solid lines) and  $\Tilde{\rho}^+(x)$ in \eqref{rhotildeasypm} (dashed lines), for $a=0.9, 4$ and $10$. The colors follow the previous figures.}
    \label{figsollasy2}
\end{figure}

It is worth commenting that the potentials \eqref{potpmasy} support other lump solutions. To obtain them, one must form the kink-antikink pair \eqref{lumpgeral} with the kink solution \eqref{soltilde}. We call these solutions $\Tilde{\phi}_l^\pm(x)$. For convenience, we take $n=-1$ to construct $\Tilde{\phi}_l^+(x)$, which has the form
\be\label{lumptildep}
\Tilde{\phi}_l^+(x) =
2\arctan\!\left(
\frac{P_1^- + P_0^- \tanh^2(x/2)}
{Q_1^- + Q_0^- \tanh^2(x/2)}
\right)-2\pi.
\ee
To obtain $\Tilde{\phi}_l^-(x)$ we consider $n=0$, so we have 
\be\label{lumptildem}
\Tilde{\phi}_l^-(x) =
2\arctan\!\left(
\frac{P_1^+ + P_0^+ \tanh^2(x/2)}
{Q_1^+ + Q_0^+ \tanh^2(x/2)}
\right).
\ee
The solution $\Tilde{\phi}_l^+(x)$ returns to the vacuum $\phi^{(-1)}_+$ and $\Tilde{\phi}_l^-(x)$ returns to the vacuum $\phi^{(1)}_-$. The choices of $n$ to construct the ``tilded'' solutions are made to facilitate their visualization in the plots. In Fig.~\ref{figsollasy1} we can see the lumps $\phi^+(x)$ in \eqref{lumpasy} and $\Tilde{\phi}^-(x)$ in Eq.~\eqref{lumptildem} and their corresponding energy densities for $\nu=0.5$ and several values of $a$. In Fig.~\ref{figsollasy2}, we display the lumps  $\phi^-(x)$ in \eqref{lumpasy} and $\Tilde{\phi}^+(x)$ in Eq.~\eqref{lumptildep} and their corresponding energy densities for $\nu=0.5$ and several values of $a$.

These solutions \emph{do not} generate another pair of potentials. In fact, $\Tilde{\phi}_l^\pm(x)$ generates $V_l^\mp(\phi)$, being the other solution that we have commented above. The energy density of the above lump is
\be\label{rhotildeasypm}
\Tilde{\rho}_l^\pm(x)=\frac{4K_\mp^2\,T^2S^4}{\left[\left(P_1^\mp+P_0^\mp T^2\right)^2+\left(Q_1^\mp+Q_0^\mp T^2\right)^2\right]^2}.
\ee
As for the lumps $\phi_l^\pm(x)$, the expression of the energy of $\Tilde{\phi}_l^\pm(x)$ is cumbersome, so we omit it here.

\section{Superposition of vacuumless kinks} \label{model4}
\begin{figure}[t!]
    \centering
    \includegraphics[width=0.75\linewidth]{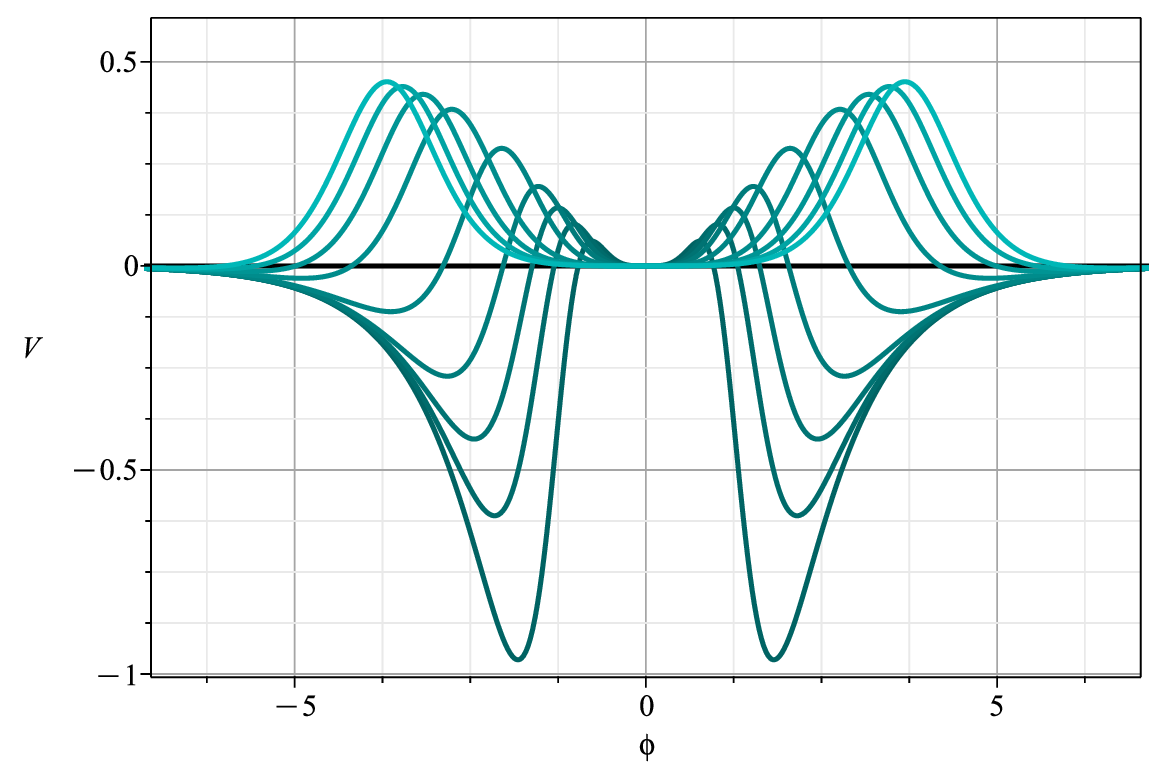}
    \caption{The potential \eqref{eq:potmodel4} depicted for $a=0.5,0.7,0.9,1.2,2,4,6,8$ and $10$. The shades of green follow the previous figures.}
    \label{fig:potvaclong}
\end{figure}

\begin{figure}[t!]
    \centering
    \includegraphics[width=0.5\linewidth]{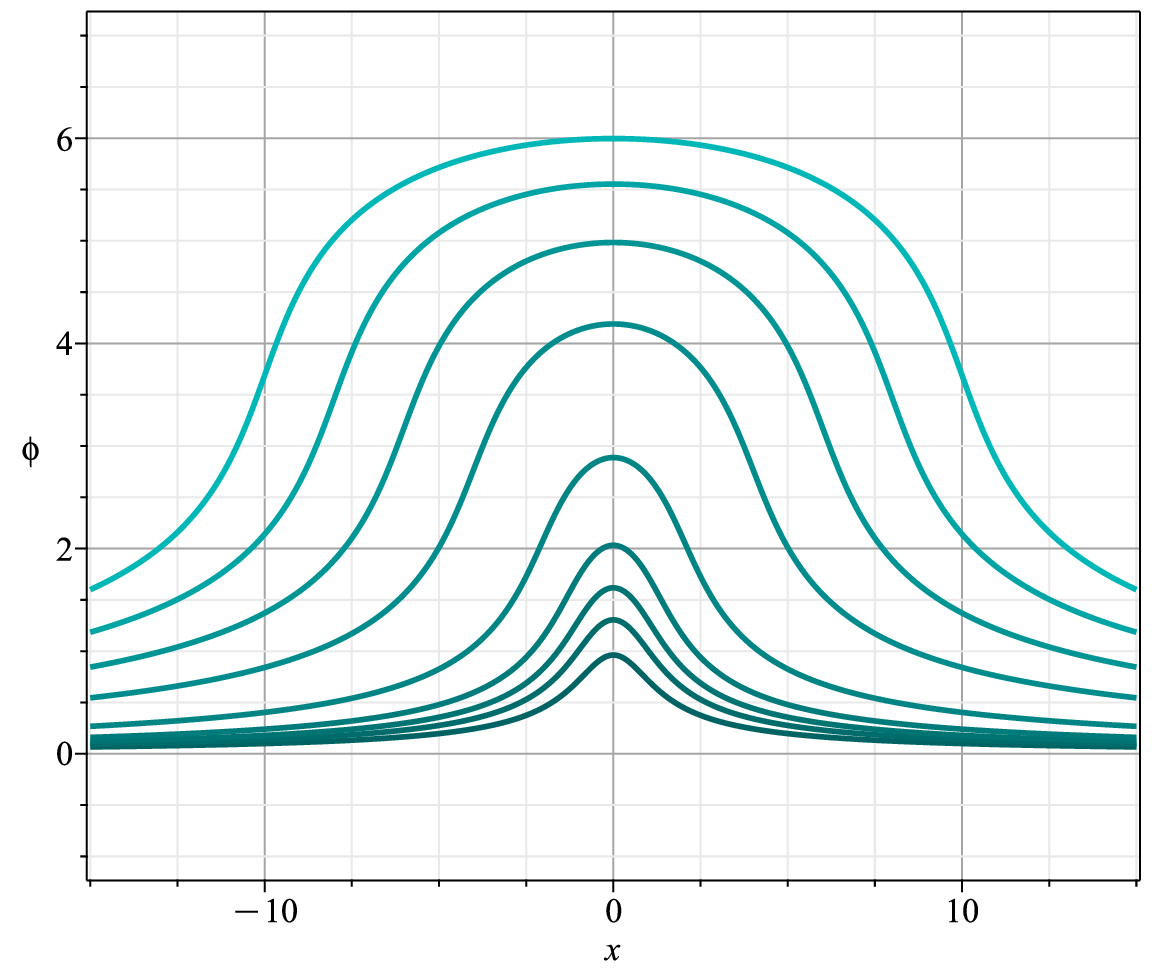}\includegraphics[width=0.5\linewidth]{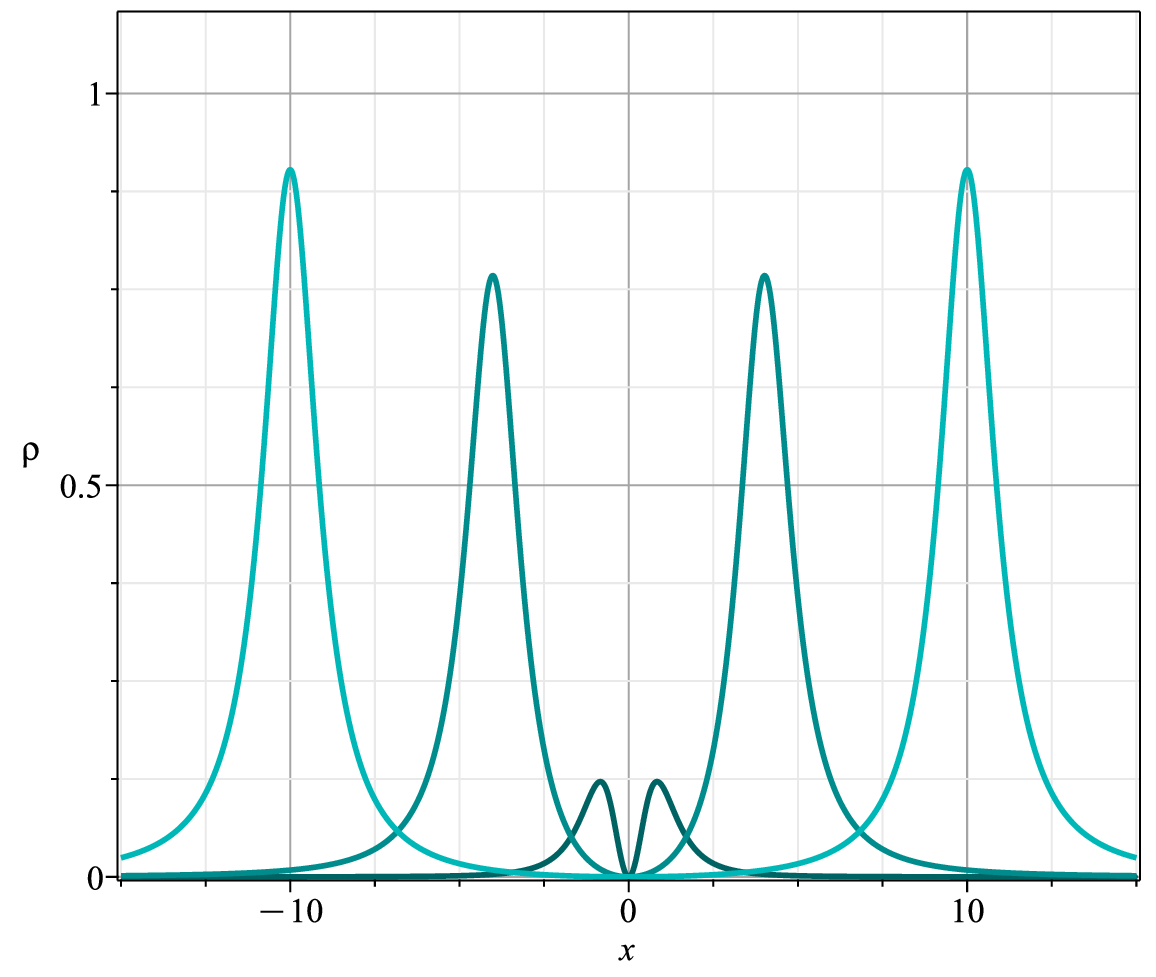}
    \caption{The solution \eqref{eq: sol model4} (left) for $a=0.5,0.7,0.9,1.2,2,4,6,8 $ and $10$, and the energy density \eqref{rholumpvac1} (right) for $a=0.5, 4$ and $10$. The colors follow the previous figures.}
    \label{figsolvaclong}
\end{figure}

We now construct a lump using the model introduced in Refs.~\cite{vac1,vac2,vac3}, defined by
\be\label{phikvac}
V_k(\phi)=\frac12\,\sech^2(\phi),\quad \phi_k = \arcsinh(x).
\ee
The energy density is $\rho_k(x) = 1/(1+x^2)$ whose integral over all space leads to $E_k=\pi$. The above solution engenders an atypical behavior, with $\phi$ spanning from $-\infty$ to $+\infty$. Therefore, the approach considered in Eqs.~\eqref{lumpgeral} is not appropriate, as it involves asymptotic values of the solution. This obstacle, however, can be surpassed by considering the lump solution $\phi(x) = \arcsinh(x+a) - \arcsinh(x-a)$, which can be written in the form
\begin{equation}\label{eq: sol model4}
    \phi_l(x) = \ln{\left(\frac{x +a +\sqrt{1 + (x+a)^2}}{x -a +\sqrt{1 +(x -a)^2}}\right)}.
\end{equation}
Even though the kink in \eqref{phikvac} diverges logarithmically, the kink-antikink pair cancels such divergence out so the lump supports power-law tails. Indeed, an asymptotic expansion of the above expression leads to $\phi_l = 2a/|x| + {\cal O}(1/x^3)$, which vanishes more slowly than the lump \eqref{lumparctan}; this is a consequence of the vacuumless character of the model \eqref{phikvac}. The potential that supports the above solution is

\begin{equation}\label{eq:potmodel4}
V(\phi)=\cfrac{2\sinh^{4}\!\left(\!\cfrac{\phi}{2}\right)\left(a^{2}-\sinh^{2}\!\left(\!\cfrac{\phi}{2}\right)\right)}{\left(a^{2}+\sinh^{4}\!\left(\!\cfrac{\phi}{2}\right)\right)^{2}}.
\end{equation}
Near the origin, it behaves as $V(\phi)\approx\frac1{8a^2}\phi^4$. The above potential possesses a minimum at $\phi=0$ and non-minimum zeros at $\phi^\pm_0 =\pm 2\arcsinh(a)$, and vanishes asymptotically. It is non-negative in the interval $|\phi|\leq2\arcsinh(a)$. In Fig.~\ref{fig:potvaclong}, we display the above potential for several values of $a$.
The lump solution \eqref{eq: sol model4} leads to the energy density
\be\label{rholumpvac1}
\begin{aligned}
	\rho_l(x) &= \frac{2(x^2+a^2+1)}{(x^2+a^2+1)^2-4a^2x^2}\\
				&- \frac{2}{\sqrt{(x^2+a^2+1)^2-4a^2x^2}}.
\end{aligned}
\ee
We can expand the above expression for large $|x|$ to get $\rho_l(x) = 4a^2/x^4 + {\cal O}(1/x^6)$. The integral of the above energy density over all space is the energy
\be
E_l = 2\pi-\frac{4}{\sqrt{a^2+1}}\,K\! \left(\frac{a}{\sqrt{a^2+1}}\right),
\ee
where $K(z)$ represents the complete elliptic integral of the first kind with argument $z$. The solution \eqref{eq: sol model4} and the energy density \eqref{rholumpvac1} can be seen in Fig.~\ref{figsolvaclong}.

\begin{figure}[t!]
    \centering
    \includegraphics[width=0.5\linewidth]{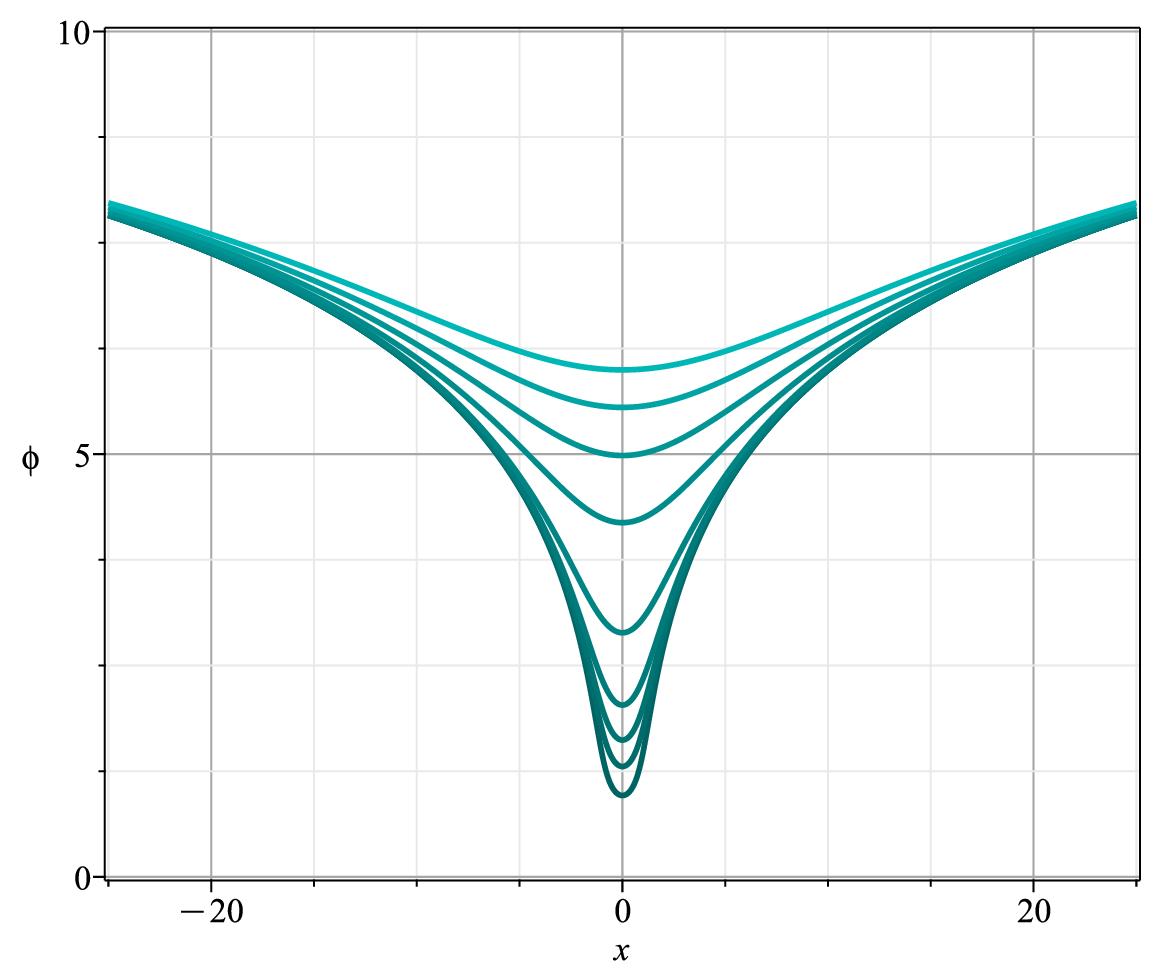}\includegraphics[width=0.5\linewidth]{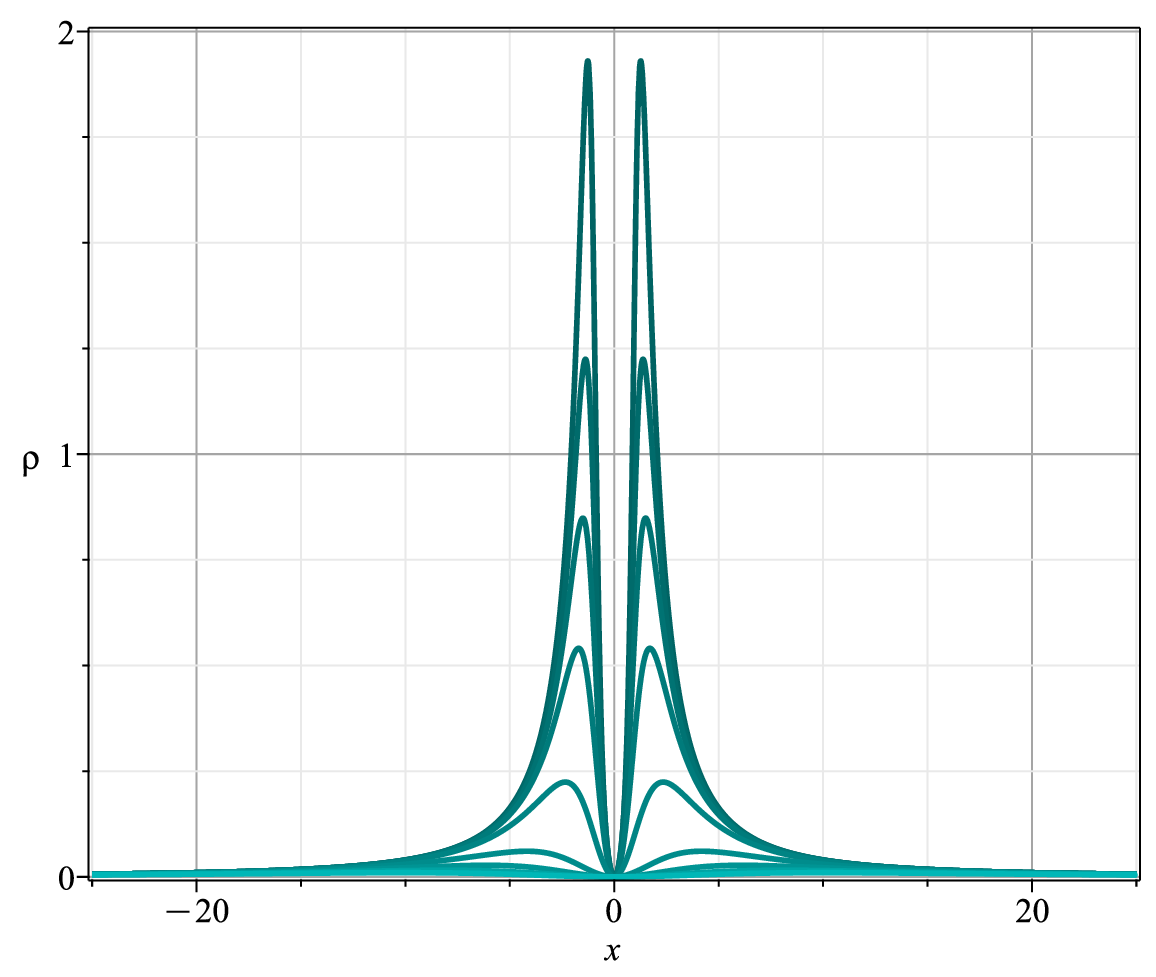}
    \caption{The solution \eqref{lumpvacuumless} (left) and the energy density \eqref{rholumpvacuumless} (right) for $a=0.5,0.7,0.9,1.2,2,4,6,8 $ and $10$. In the right panel, the lines become indistinguishable for $a\geq6$.}
    \label{figsolvacvac}
\end{figure}

The potential \eqref{eq:potmodel4} has an interesting feature: it crosses zero, becomes negative and goes asymptotically to zero via negative values. This means that we can change its sign to obtain an unprecedented non-topological solution that has infinite amplitude, i.e., it diverges asymptotically after passing through its return point. Due to this feature, we call it \emph{vacuumless} lump. We then consider
\be\label{vvacinvert}
V(\phi)=\cfrac{2\sinh^{4}\!\left(\!\cfrac{\phi}{2}\right)\left(\sinh^{2}\!\left(\!\cfrac{\phi}{2}\right)-a^2\right)}{\left(a^{2}+\sinh^{4}\!\left(\!\cfrac{\phi}{2}\right)\right)^{2}}.
\ee
\begin{figure}[t!]
    \centering
    \includegraphics[width=0.75\linewidth]{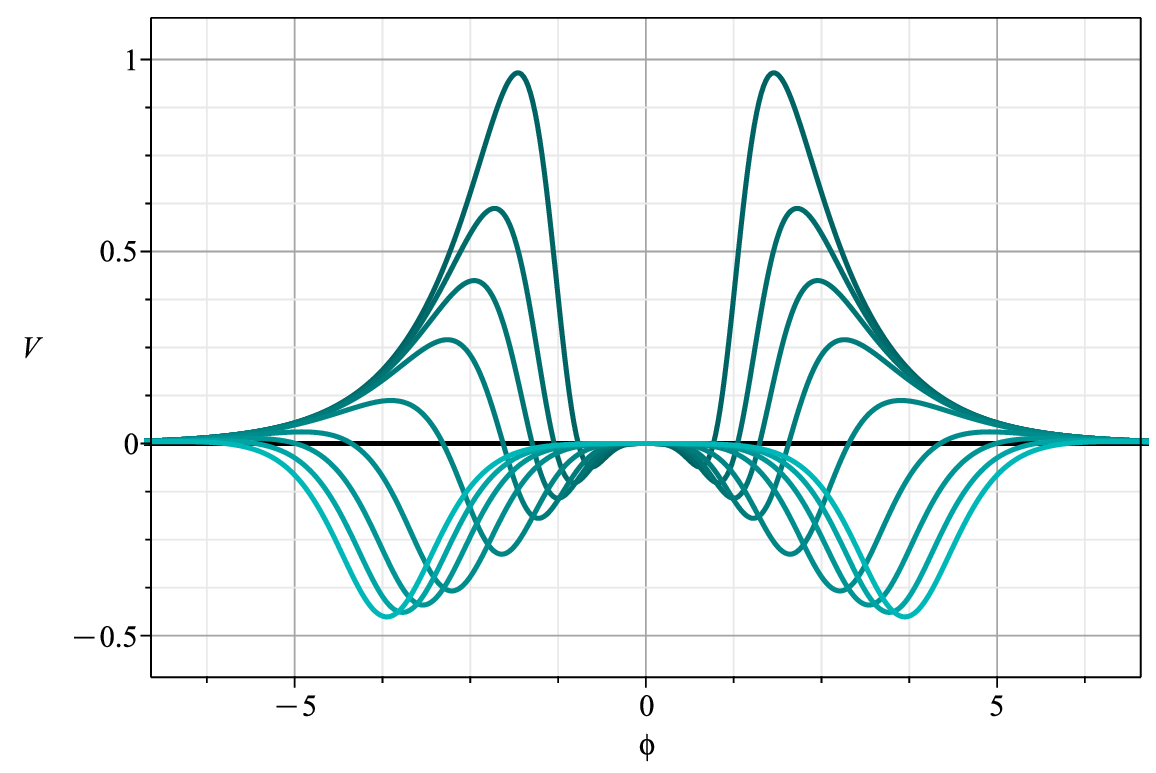}
    \caption{The potential \eqref{vvacinvert} depicted for $a=0.5,0.7,0.9,1.2,2,4,6,8$ and $10$. The shades of green follow the previous figures.}
    \label{fig:potvacvac}
\end{figure}
We display this potential in Fig.~\ref{fig:potvacvac}. It is non-negative in the interval $|\phi|\in[2\arcsinh(a),\infty)$. This property brings the following solution to light:
\be\label{lumpvacuumless}
\phi_l(x)\!=\!2\arcsinh\sqrt{\frac{x^2+a^2-1+\sqrt{(x^2+a^2-1)^2+4a^2}}{2}}.
\ee
The asymptotic expansion shows that the above expression diverges logarithmically as $\phi_l = 2\ln(2|x|) + {\cal O}(1/|x|)$. This occurs due to the vacuumless-like behavior of the model. The energy density of \eqref{lumpvacuumless} is
\be\label{rholumpvacuumless}
\rho_l(x)=\frac{2\left[\sqrt{(x^2+a^2-1)^2+4a^2}+x^2-a^2-1\right]}{(x^2+a^2-1)^2+4a^2}
\ee
and the energy is
\be
E_l = \frac{4}{\sqrt{a^2+1}}\,K\! \left(\frac{1}{\sqrt{a^2+1}}\right).
\ee
Therefore, the energy is finite, despite the solution being divergent. This occurs because of the logarithmic behavior of the lump solution, which \emph{does not} delocalize the energy density. Verily, the asymptotic behavior of \eqref{rholumpvacuumless} is
\be
\rho_l(x)= \frac4{x^2} + {\cal O}\left(\frac1{x^4}\right).
\ee
This shows that the energy density is indeed localized, ensuring finite energy. The power-law tail of the above $\rho_l(x)$ falls off more slowly than \eqref{rholumpvac1} ($\rho_l(x)\propto1/x^4$) which falls off more slowly than \eqref{rhosglong} ($\rho_l(x)\propto1/x^6$). The behavior of the solution \eqref{lumpvacuumless} and its energy density \eqref{rholumpvacuumless} can be seen in Fig.~\ref{figsolvacvac}.
\section{Final remarks}\label{secfinalremarks}
In this work, we have investigated how to construct lump solutions as the superposition of kinks that form a kink-antikink pair. To do so, we have considered two possibilities, related to the downshifting of each minimum of the potential connected by the parent kink; see Eqs.~\eqref{lumpgeral}. The potential associated with each lump can be obtained via the first-order equation \eqref{fo}. This, however, requires us to find the inverse of the lump solution $\phi_l(x)$, which is not always feasible analytically. If the parent kink is symmetric, the potential that originates the lump $\phi^+(x)$ is equivalent to the one for $\phi^-(x)$, so a single model can be constructed. On the other hand, asymmetric kinks may lead to two distinct models. In both cases, nevertheless, the lump inherits the tails of the kink used in the superposition. Interestingly, even though the lump solution is given by a simple sum of a kink and antikink equidistant from the origin, the energy density of the lump \emph{is not} a linear combination of their energy densities; a term of interaction appears, showing how the mixing of the substructures enters in our procedure.

Our method allows us to obtain novel analytical lump solutions. We have first tested its robustness by considering toy and $\phi^4$ parent models. Then, we have considered sine-Gordon, double sine-Gordon and long-range sine-Gordon kinks to originate new lumps, obtaining their corresponding potential and calculating the physical properties of interest, such as the critical points of the potentials, energy density and energy. All the parent kinks in Sec.~\ref{secsymmetric} are symmetric. Since the presence of asymmetry may lead to distinct features, we have investigated this issue in Sec.~\ref{secasymmetric}. There, we have first introduced a novel asymmetric kink solution; see Sec.~\ref{secasymmetrickinks}. This was done to obtain analytical results in the corresponding lump section. We have then discussed the main features of these kinks. In Sec.~\ref{seclumpsasykinks}, we have used these novel asymmetric kinks to build lumps with our procedure. As we have commented before, this case is richer, with two distinct potentials arising in the process.

The last model that we have investigated is the one with lumps emerging from vacuumless kinks as their parents. As we have discussed in Sec.~\ref{model4}, this case is special because the kink spans from $-\infty$ to $\infty$. Therefore, the superposition \eqref{lumpgeral} \emph{must} be modified. To overcome this difficulty, we simply remove the asymptotic value of the field from the superposition, giving rise to a long-range lump. We also construct the potential and show that the energy density is localized, leading to finite energy. The lump model that arises from the superposition of vacuumless kinks engenders an interesting feature: by changing the sign of the lump potential, we were able to obtain a vacuumless lump, an unprecedented solution with localized energy density and finite energy that is not obtained via superposition.

Since our procedure shows how to construct a lump using a kink-antikink pair, a natural question that arises is whether any lump can be obtained using our technique. Before ending our investigation, let us address this issue. We consider a one-parameter family of lump profiles $\phi_l(x;a)$, where $a$ is a parameter that controls the separation between the two substructures. If this family can be obtained from the standard kink-antikink superposition, then there must exist a kink profile $\phi_k(x)$ and a constant $C$ such that $\phi_l(x;a)=\phi_k(a+x)+\phi_k(a-x)-C$. A necessary condition follows from the second and higher derivatives with respect to $a$ and $x$. Indeed, any lump generated by the superposition must obey
\be
\frac{\partial^{2\ell} \phi_l(x;a)}{\partial a^{2\ell}}
-
\frac{\partial^{2\ell} \phi_l(x;a)}{\partial x^{2\ell}}
=0,
\ee
for all positive integer $\ell$. This provides a simple diagnostic test: if the above equality is not satisfied, then the lump solution $\phi_l(x;a)$ cannot be obtained from a standard kink-antikink superposition with an $a$-independent parent kink. Indeed, all the lump solutions in this manuscript, except \eqref{lumpvacuumless}, satisfy the above equality. This confirms that this specific solution \emph{cannot} be obtained with our procedure.

As future perspectives, since the lumps are unstable and spontaneously decay over time, one may investigate how the different asymptotic behaviors, such as the exponential and the three types of power-law tails, affect the decay of the large lump (large $a$), to verify whether their collisions mimic the scattering behavior found in Ref.~\cite{lumpscatt}. We can also think of adding fermions into the system, coupling them to the scalar field in a way similar to the cases already considered in Refs. \cite{MC,MB}, concerning the formation of Fermi balls; see also some more recent work on the use of Fermi balls \cite{Fb1,Fb2} . The issue here is to investigate how fermions can bind into the localized scalar configurations described in the present work to make them contribute to stabilizing the composite structure as a novel possible dark matter candidate. It is also of interest to recall the recent review \cite{RPP}, which includes an updated account of some properties of several non-topological configurations that have significant applications in cosmology and particle physics. Another issue of interest concerns the possibility of extending the above procedure to the case of bright and dark solitons in fibers and in Bose-Einstein condensates. These and other related issues are presently under consideration, and we hope to report on them in the near future.

\acknowledgments{
{This work is supported by the Conselho Nacional de Desenvolvimento Científico e Tecnológico (CNPq), grants 402830/2023-7 (DB, MAM, and RM), 303469/2019-6 (DB), 303875/2026-7 (MAM) and 304344/2025-7 (RM), and by the Coordenação de Aperfeiçoamento de Pessoal de Nível Superior (CAPES), grant 88887.899549/2023-00 (IB).}}

\end{document}